\documentclass[a4paper,12pt,twocolumn,journal]{IEEEtran}
\usepackage{graphicx,xcolor}
\usepackage{xspace}
\usepackage{calc}
\usepackage{tabularx}
\usepackage{amsmath}
\usepackage{braket}
\usepackage{relsize}    
\usepackage{mathabx}  
\usepackage{bbm}		

\DeclareMathOperator{\mean}{mean}
\DeclareMathOperator{\median}{median}
\DeclareMathOperator{\modedist}{mode}  
\DeclareMathOperator{\variance}{variance}
\DeclareMathOperator{\Prob}{Prob}
\DeclareMathOperator{\sech}{sech}
\newcommand{\dd}{\mathop{}\!\mathrm{d}}

\newtheorem{theorem}{Theorem}
\newtheorem{lemma}[theorem]{Lemma}

\newtheorem{corollary}[theorem]{Corollary}

\newcommand{\rvN}{{\cal N}}
\newcommand{\rvLN}{{\cal LN}}

\newcommand{\Z}{{\mathbbm Z}}
\newcommand{\Zp}{\Z_{+}}
\newcommand{\R}{{\mathbbm R}}
\newcommand{\Rp}{\R_{+}}
\newcommand{\fN}[1]{f_{\cal N}\!\left(#1\right)}
\newcommand{\erf}[1]{\text{erf}\!\left(#1\right)}

\newcommand{\fa}[1]{f_{\text{a}}\!\left(#1\right)}
\newcommand{\Fa}[1]{F_{\text{a}}\!\left(#1\right)}
\newcommand{\cFa}[1]{\bar{F}_{\text{a}}\!\left(#1\right)}
\newcommand{\supporta}{S_{\text{a}}}

\newcommand{\cFxK}[1]{\bar{F}_{\text{a}_K}\!\left(#1\right)}

\newcommand{\fxk}[1]{f_{\text{x}_k}\!\left(#1\right)}
\newcommand{\fxkp}[1]{f_{\text{x}_{k+1}}\!\left(#1\right)}
\newcommand{\fxK}[1]{f_{\text{x}_K}\!\left(#1\right)}

\newcommand{\falpha}[1]{f_{\alpha}\!\left(#1\right)}

\newcommand{\cFalpha}[1]{\bar{F}_{\alpha}\!\left(#1\right)}
\newcommand{\supportalpha}{S_{\alpha}}

\newcommand{\FalphaK}[1]{F_{\alpha_K}\!\left(#1\right)}

\newcommand{\falphak}[1]{f_{\alpha_k}\!\left(#1\right)}
\newcommand{\falphaK}[1]{f_{\alpha_K}\!\left(#1\right)}

\newcommand{\fdelta}[1]{f_{\delta}\!\left(#1\right)}

\newcommand{\fzetak}[1]{f_{\zeta_{k}}\!\left(#1\right)}
\newcommand{\fzetakp}[1]{f_{\zeta_{k+1}}\!\left(#1\right)}
\newcommand{\fzetaK}[1]{f_{\zeta_K}\!\left(#1\right)}

\newcommand{\xbnd}{x_{\text{bnd}}}

\newcommand{\e}[1]{\text{e}^{#1}}

	\newcommand{\be}{\begin{equation}}
	\newcommand{\ee}{\end{equation}} 
	\newcommand{\bu}{a}
	
	\newcommand{\dexpec}[1]{{\cal E} \left[ \rule{0em}{.9em} #1 \right]}
    	\newcommand{\req}[1]{(\ref{#1.eq})}
	
\newcommand{\Matlab}{{\sc Matlab}}

\newlength{\figwidth}
\definecolor{gold}{rgb}{0.85,0.64,0.13}
\definecolor{cyan4}{rgb}{0,0.55,0.55}

\title{Stochasticity in Feedback Loops\\*[0.25em]
\Large Great Expectations and Guaranteed Ruin}
\author{Roy S.\ Smith and Bassam Bamieh%
\thanks{Roy Smith (corresponding author) is with the Automatic
Control Laboratory, ETH Z\"{u}rich, Switzerland.}
\thanks{Bassam Bamieh is with the Mechanical Engr.\ Dept., University
of California at Santa Barbara, CA, USA.}
}

\begin{document}

\maketitle

\section{Introduction}

Stochastic components in a feedback loop introduce state behaviours that are fundamentally
different from those observed in a deterministic system.  The effect of injecting a stochastic signal
additively in linear feedback systems can viewed as the addition of filtered stochastic
noise.  If the stochastic signal enters the feedback loop in a multiplicative manner a
much richer set of state behaviours emerges.  These phenomena are investigated for
the simplest possible system;  a multiplicative noise in a scalar, integrating feedback 
loop.   The same dynamics arise when considering a first order system in feedback 
with a stochastic gain.   Dynamics of this form arise naturally in a number of domains,
including compound investment in finance, chemical reaction dynamics, population dynamics, control over
lossy communication channels, and adaptive control, to mention a few.   Understanding
the nature of such dynamics in a simple system is a precursor to recognising them in more
complex stochastic dynamical systems.

Some of the results presented have appeared in other research domains in the past, 
but are not widely known with in the control systems community.  The presentation and
proof of the results depends only on reasonably well known statistical results.   This paper
draws upon  and augments such results to study the stability of a stochastic feedback loop.  

The earliest formulation of the problem we consider was formulated by Kalman~\cite{kalman:1962a}
and reports on results from his Ph.D.\ thesis.  A discrete-time control design problem is posed,
with the open-loop system being a discrete-time difference equation model of the form,
\begin{equation}
\label{e:kalman}
x_{k+1} \; = \; A_{k+1}x_{k} + B_{k+1} u_{k},
\end{equation}
where $A_{k+1}$ and $B_{k+1}$ are elements of stationary, mutually-independent distributions.
The design criteria is mean-square stability which essentially means that  from any given
$x_0$ the variance is finite and decays to zero as $k\longrightarrow\infty$.    This problem was
also considered and characterised in the frequency domain for both continuous- and discrete-time
systems by Willems and Blankenship in~\cite{willems:1971a}.

Mean-square
stability conditions are appealing as they can be formulated in the multivariable case and
relate directly to covariance matrices~\cite{willems:1973a}.
Furthermore they lead to convex optimisation problems for analysis and, in some cases,
controller design (see for example Boyd {\it et al.}~\cite{boyd:1994a}).  The
disadvantage, that will become obvious here, is that mean-square stability is a very strong
form of stability.  In many applications something weaker might be preferable.

Adaptive control research began in the 1960s and motivated the study of feedback systems with
stochastically varying parameters.  The work by {\AA}str\"{o}m in~\cite{astrom:1965a}
corresponds most closely to the approach taken here in that it characterises the
distributions that result in such systems. 
A continuous-time setting was used in~\cite{astrom:1965a} and much of the earlier work, which
considers stochastic differential equations of the form,
\[
dx \; = \; x \, dw_1 + dw_2,
\]
where $dw_1$ and $dw_2$ are Wiener processes.  Some of the characteristics of
the limiting distributions in~\cite{astrom:1965a} are also evident in the distributions 
arising in this paper.  In contrast, the work here considers stochastic difference equations
where the multiplicative term can be drawn from a wide range of possible distributions.

A continuous-time setting was also used by Blankenship~\cite{blankenship:1977a} with
the system model,
\[
\dot{x}(t) = A(t) x(t), \quad x(t_0) = x_0 \in\R^n,
\]
where the elements of $A(t)$ are stochastic stationary processes with certain 
continuity properties.  The results use differential equation solution bounds to
give sufficient conditions under which,
\begin{equation}
\label{e:limzero}
\Prob\left\{ \lim_{t\longrightarrow\infty} |x(t)| = 0 \right\} = 1.
\end{equation}
This is a significantly weaker form of stability and in the scalar framework of this
paper is equivalent to the stability of the median of $|x(t)|$.

This paper will make the case that in many applications the stability of the median is an
important practical concept.   Interestingly a similar case has also been made in the
domain of gambling strategies~\cite{ethier:2004a} where it was observed that
proportional betting---a multiplicative strategy analogous to the stochastic feedback
configuration---optimises the median of the gambler's fortune.    We will see that gamblers
care about the median as it characterises their probability of making a profit.  In contrast the
gambling house cares about the mean as it characterises their risk.

From a probability theory point of view the work presented here can be considered
as an application of renewal theory.   The problem of determining properties 
of the limiting distribution of the matrix evolution,
\begin{equation}
x_{k+1} \; = \; A_{k} x_{k} + q_{k}, \quad x\in\R^n, \quad k \geq 0, \label{e:kesten}
\end{equation}
where $A_k$ is a random positive matrix and $q_k$ a random vector, has been studied
by Kesten~\cite{kesten:1973a}.  The conditions under which there exists a limiting distribution,
$f_{x_{\infty}}(x)$,  are given and are essentially a generalisation of the 
median stability results presented in this paper.  In the scalar case~\cite{kesten:1973a}
showed that $f_{x_{\infty}}(x)$ can be heavy-tailed, even if the distributions of $A_k$ and
$q_k$ are relatively light-tailed.
Kesten's work was extended in work by Goldie~\cite{goldie:1991a} where a range of
similar recursions were shown to also 
give power-law distribution tails.  The limiting distribution, when it exists, was derived
by Brandt~\cite{brandt:1986a}.  The existence conditions are essentially equivalent to that for
the stability of the median derived in our work.  
Work by de Saporta~\cite{saporta:2005a} describes an
interesting variation on the recursion of~(\ref{e:kesten}), by considering the limiting
distribution in the case where $A_k$ comes from a Markov chain.  Much more on the stochastic
stability of Markov chains can found in the comprehensive text of Meyn and Tweedie~\cite{meyn:2009a}.

One of the applications considered in this paper is the stabilisation of an unknown system via
stochastic feedback.  This has also been considered by Milisavljevi\'{c} and
Verriest~\cite{milisavljevic:1997a} and they provide a stability condition which is an application
of our results on median stability. 

The growth in research interest and application of networked control systems has introduced 
another application of this theory.  Sinopoli {\it et al.}~\cite{sinopoli:2004a} showed that 
Kalman filters with intermittent observations can lose mean-square stability once the probability
of missing a packet reaches a threshold value.  The focus on mean-square stability is natural in
the Kalman filtering case as the construction of the time-varying Kalman filter requires a well-defined
covariance matrix evolution.  In the case of a static Kalman gain, the
evolution of the estimation error is of the form given in (\ref{e:kesten}).   An analagous result on 
stabilisation over fading channels was shown by Elia~\cite{elia:2005a}.   Elia also
observed  the  emergence of heavy-tailed distributions in networked control systems in the 
case where mean-square stability is lost~\cite{Elia:2006a} and provided a mathematical
characterisation of this behaviour in~\cite{wang:2012a}.  Work by Mo and Sinopoli~\cite{mo:2012a}
extended the packet loss model and provided bounds on the tail of the error
distribution. Dey and Schenato~\cite{dey:2018a} observe the distinction between the instability 
of the second moment and the conditions required for the existence of a limiting 
power-law distribution.  As also observed in our work, this is the distinction between median stability and
variance stability.

The adaptive control application that provided motivation for analysis of these systems in the 1960s 
has recently received renewed attention.  Rantzer~\cite{rantzer:2018b} considers a single parameter
case and examines the stability of various moments.  Concentration bounds on the distribution of the
parameter error are derived.

Our paper focuses on the scalar discrete-time case given in~(\ref{e:kesten}), without the
random exogenous input $q_k$, and shows that even though the distribution of $f_{x_{\infty}}(x)$
might be heavy-tailed it is still possible that (\ref{e:limzero}) holds;
the state $x_k$ decays to zero with probability one.  By focusing on the scalar case we are able to
derive and calculate the distributions as a function of the time index, $k$, and give explicit conditions
for the stability of the median, mean, and variance of the state.  The results given are not
unexpected in light of the prior work outlined above, but the explicit characterisation of stability conditions, and
the calculation of the distributions involved at each time step, provides insight into the
manner in which the instabilities are manifested.

\subsection{Notation}

The notation $a \sim \fa{a}$ is used to denote that the random variable $a$ is drawn from 
a distribution with probability density function $\fa{a}$.   The cumulative distribution function is denoted by
$\Fa{a}$ and the complementary cumulative distribution by $\cFa{a}$ ($ = 1 - \Fa{a}$). 
The expected value of $a$ is denoted by $\dexpec{a} = \mu_a$.
The normal distribution of mean $\mu$ and variance $\sigma^2$ is denoted by $\rvN(\mu,\sigma^2)$,
and the lognormal distribution by $\rvLN(\mu,\sigma^2)$.

The set of (non-negative) integers is denoted by ($\Zp$) $\Z$; and the reals by ($\Rp$) $\R$.

\section{Problem description}

The plant is a first order system with the scalar state $x_k$ evolving with dynamics given
by
\begin{equation}
x_{k+1} \: = \: a_{k}\,x_{k}, \quad k = 0,1, \dots \label{e:xdyn}
\end{equation}
where $a_k$ are independent random variables drawn from a distribution
$\fa{a}$ with mean $\mu_a$ and variance $\sigma_a^2$.  
The distribution $\fa{a}$ is assumed to have support only on $a > 0$ and 
$x_0$ is assumed to be strictly positive.

\begin{figure}
\centering
\begin{tabular}{c}
\includegraphics{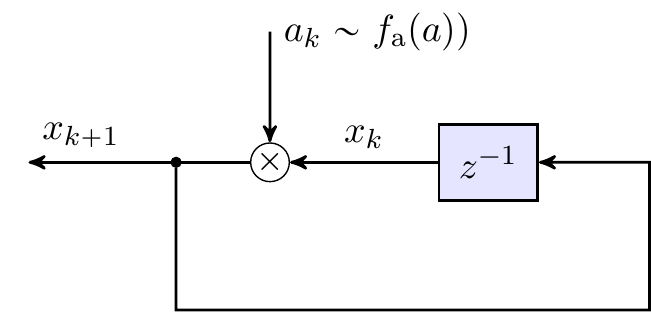}\\
\noalign{\bigskip}
\includegraphics{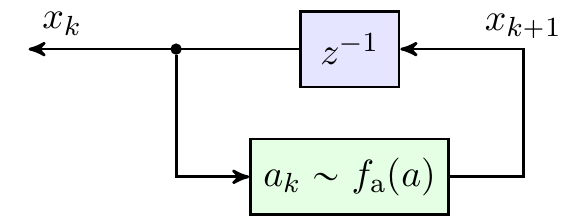}
\end{tabular}
\caption{\label{f:config}Process dynamics.  The upper system generating the 
stochastic dynamics is shown as a multiplicative noise signal in a feedback loop.
The equivalent system below is a feedback loop with a stochastic gain.
}
\end{figure}

The dynamics in~(\ref{e:xdyn}) can be viewed as multiplicative noise, $a_k$, 
entering a feedback loop.  An alternative view is that of a stochastic feedback 
gain.  Both interpretations are illustrated in Figure~\ref{f:config}.  Both involve 
a feedback loop around a delay.

This is the simplest case of the type of processes described above.  As we are 
considering only a scalar state and uncorrelated $a_k$ it is too simplistic for many 
real processes of this type.   However, it is a prototypical case and illustrates some of the 
phenomena that may arise in more complex systems.   Understanding the stability 
characteristics of this system is a precursor to understanding those for more complex
systems.

\subsection{Stability}

At the time index $K$ the state $x_K$ is given by
\[
x_K = \prod_{k=0}^{K-1} a_k\,x_0,
\]
and as $a_k \sim \fa{a}$ the state $x_K$ is also a stochastic
variable with a probability density function we denote by $\fxK{x}$.   We are 
interested in the properties of this distribution as  $K\longrightarrow\infty$.  
More specifically we will derive conditions for the following three notions of 
stability.

The system is defined as \emph{median stable} if and only if
\[
\lim_{K\longrightarrow\infty}\, \median(x_K) \, = \, 0.
\]
The system is defined as \emph{mean stable} if and only if
\[
\lim_{K\longrightarrow\infty}\, \dexpec{x_K} \, = \, 0,
\]
and the system is defined as \emph{variance stable} if and only if
\[
\lim_{K\longrightarrow\infty}\, \dexpec{(x_K - E\{x_K\})^2} \, = \, 0.
\]

The approach taken will involve analysing the system in terms of the 
probability density functions of the logarithmic variables,
\begin{equation}
\zeta_{k} := g(x_k) = \ln(x_k) \label{e:zetadefn}
\end{equation}
and
\begin{equation}
\alpha_k := g(a_k) = \ln(a_k) \label{e:alphadefn}.
\end{equation}
The function $g$ is defined here for convenience in subsequent derivations.

We will assume that $x_0 > 0$ is given and so $x_K \geq 0$ for all $K\in\Zp$.
For simplicity in the following we will assume without loss of generality that 
$x_0 = 1$.
The evolution of the dynamics in~(\ref{e:xdyn}) now becomes
\begin{equation}
\zeta_{k+1} = \zeta_{k} + \alpha_k. \label{e:zetadyn}
\end{equation}
This allows us to express $x_K$ as,
\[
x_K \; = \;  \e{\zeta_K}
\]
To illustrate the way in which stability results will be derived we can
examine this evolution for $K$ timesteps.  Under our assumption that $x_0 = 1$,
\[
\dexpec{\zeta_K}  = \dexpec{ \sum_{k=0}^{K-1}\alpha_k } 
		    = K \dexpec{ \alpha } = K\mu_{\alpha},
\]
if the distributions $\falphak{\alpha}$ are identically distributed.

It is tempting to say that if $\mu_{\alpha} < 0$ then $\dexpec{x_K} < 1$.   This is 
not true as the results in 
the following sections will make clear.
What will turn out to be true is that for distributions where the distribution 
of $\ln(a)$ satisfies
certain moment assumptions,
\[
\dexpec{ \alpha } \, < \, 0 \quad \iff \quad
\lim_{K\longrightarrow \infty} \median(x_K) \, = \, 0.
\]

\subsection{Commutative variable relationships}

\begin{figure*}
\centering
\includegraphics{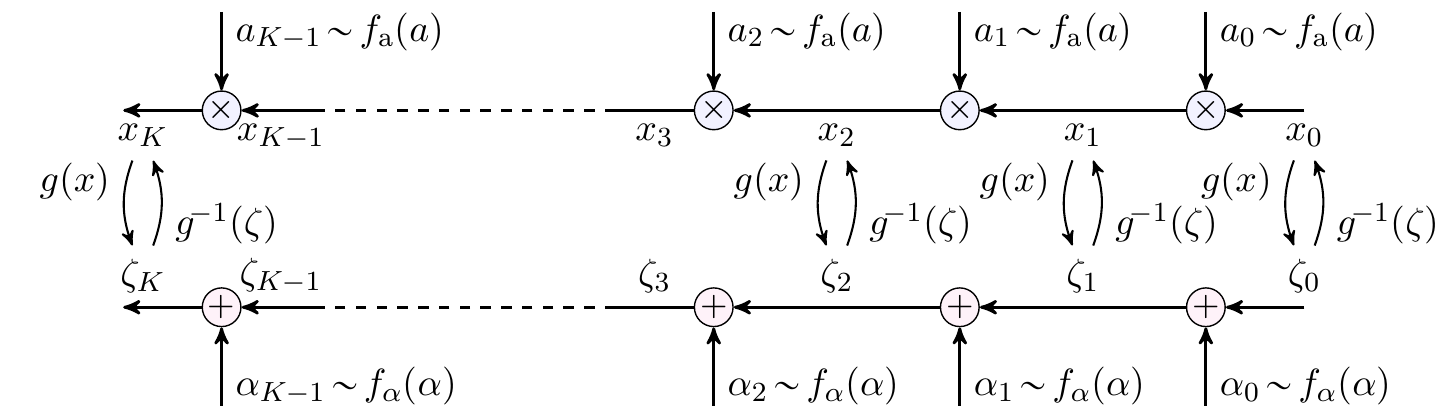}
\caption{\label{f:commutative} State variable evolution.  The mapping between the $x$ and
$\zeta$ variables is given by $\zeta_i = g(x_i) = \ln(x_i)$, and the inverse mapping is
$x_i = g^{-1}(\zeta_i) = \e{\zeta_i}.$ }
\end{figure*}

The system will be studied in terms of the probability distributions of
both the $x_k$, $a_k$ variables and their logarithmic versions,
$\zeta_k$, $\alpha_k$.     The logarithmic/exponential relationship between these
variables means that one can map the distributions from one set of 
variables to the other.  Figure~\ref{f:commutative} gives a commutative 
diagram of these relationships as they evolve over the time index.

The mappings to logarithmic variables in~(\ref{e:zetadefn}) and~(\ref{e:alphadefn}) maps
the corresponding distributions.  This is described for the $\alpha$ variable but also
applies to the $x_k$ variables.   
Suppose that $a$ has a probability distribution $\fa{a}$ defined
on a support $\supporta \subseteq \Rp$,
\[
\supporta = \left\{ a \, \mid \, \fa{a} > 0 \right\}.
\]
If $g(a)$ is monotonically
increasing and invertible on $\supporta$ then the probability distribution of $\alpha$ is
given by,
\begin{equation}
\falpha{\alpha} = \left\{ \begin{array}{ll}
    \fa{g^{-1}(\alpha)} \displaystyle \left| \frac{\dd\, g^{-1}(\alpha)}{\dd\alpha} \right|, \: & \text{if } \alpha \in \supportalpha, \\
          0,  & \text{if } \alpha \notin \supportalpha.
          \end{array}\right. \label{e:vdist2alphadist}
\end{equation}
and has support,
\[
\supportalpha = \left\{ \alpha = g(a) \, \mid \, a \in \supporta \right\}.
\]

In our case 
\[
g(a) = \ln(a), \quad  g^{-1}(\alpha) = \e{\alpha},
\]
and
\[
\left| \frac{\dd\, g^{-1}(\alpha)}{\dd\alpha} \right| = \left| \e{\alpha} \right| = \e{\alpha}.
\]

We are interested in the distribution of $x_K$ as $K$ increases and, as we can see from
Figure~\ref{f:commutative}, there are several ways of calculating this distribution.  One can
directly consider the evolution of the variable,
\[
x_{k+1}\: = \: a_k \,x_{k},
\]
where $x_{k}$ has density $\fxk{x}$ and $a_k$ has density $\fa{a}$.  This can be
calculated as,
\[
\fxkp{x} = \int_{-\infty}^{\infty} \fxk{\xi} \fa{x/\xi} \frac{1}{| \xi |} \dd \xi.
\]
The $\fxkp{x}$ distribution can also be obtained by first transforming $x_k$ to 
$\zeta_k$ using the mapping in~(\ref{e:vdist2alphadist}),
\[
\fzetak{\zeta} = \fxk{\e{\zeta}} \e{\zeta}.
\]
The $\zeta$ dynamics are simply additive (see~(\ref{e:zetadyn}))
and so the $\zeta_{k+1}$ distribution is given by the convolution,
\[
\fzetakp{\zeta} = \int_{-\infty}^{\infty} \fzetak{\xi} \fa{\zeta - \xi} \dd \xi.
\]
The $\fxkp{x}$ distribution is then given by the inverse of the
mapping in~(\ref{e:vdist2alphadist}),
\[
\fxkp{x} = \fzetakp{\ln(x)} \frac{1}{x}, \quad x > 0.
\]

\subsection{Lognormal distributions}
\label{s:lognormal}

The case where $\fa{a}$ is a lognormal distribution is in some sense generic.  If $\fa{a}$ is 
a lognormal distribution, then $\falpha{\alpha}$ is normal. Then the $\zeta_K$ distribution is a sum of
normal distributions and is also normal.   Equivalently the distribution of $x_K$ is lognormal
for all $K$.  In other words,  lognormal distributions are closed under multiplication of the random
variables.

The central limit theorem implies that even if the $\falpha{\alpha}$ distribution is not normal, the scaled
distribution of $\zeta_K$ tends to a normal distribution.  One therefore also expects that the 
scaled $x_K$ distribution tends to a lognormal distribution.   We present some of the
properties of lognormal distributions for later use.  All of the results here can be found
in~\cite{aitchison:1957a}.

The lognormal distribution can be defined by considering  $a$ as given by,
\[
a = \e{\alpha} \quad\text{where}\quad \alpha \; \sim \; \rvN(\mu_{\alpha},\sigma^2_{\alpha}).
\]
The parameters $\mu_{\alpha}$ and $\sigma_{\alpha}$ are known as the location and scale parameters respectively,
but here we simply refer to them as the mean and standard deviation in the $\alpha$-space.
This distribution definition ensures that $\supporta = \Rp$ and so fits the assumptions
of our problem.

From~(\ref{e:zetadyn}) it is clear that $\zeta_K$ is the sum of $K$ random variables, $\alpha_k$,
with each $\alpha_k \sim \rvN(\mu_{\alpha},\sigma^2_{\alpha})$.  The sum of normally distributed variable is also normally
distributed and
\[
\zeta_K \, \sim \, \rvN( K\mu_{\alpha},K\sigma^2_{\alpha}).
\]
This gives a closed-form expression for the distribution of $\zeta_K$,
\begin{equation}
\fzetaK{\zeta} = \fN{\alpha,K\mu_{\alpha},K\sigma^2_{\alpha}}, \label{e:zetadist}
\end{equation}
where
\[
\fN{x,\mu,\sigma^2} = \frac{1}{\sqrt{2\pi\sigma^2}} \e{-(x-\mu)^2/2\sigma^2}.
\]
Closed-form expressions relate the mean and variance of $a_k$ to the mean
and variance of $\alpha_k$~\cite{finney:1941a,shellard:1952a}.
\begin{IEEEeqnarray}{rCl}
\mu_{\alpha} &  =  &  \ln \left( \frac{\mu_a}{\sqrt{1 + \frac{\sigma_a^2}{\mu_a^2}}} \right) \label{e:mualpha} \\*
\noalign{\medskip}
\sigma_{\alpha}^2 &  = &  \ln \left( 1 + \frac{\sigma_a^2}{\mu_a^2} \right). \label{e:varalpha}
\end{IEEEeqnarray}
The inverse mapping is given by,
\begin{IEEEeqnarray}{rCl}
\mu_a &  = &  \e{\mu_{\alpha} + \sigma_{\alpha}^2/2} \label{e:muv} \\*
\noalign{\medskip}
\sigma_a^2 &  = &  \left( \e{\sigma_{\alpha}^2} - 1 \right) \left( \e{2\mu_{\alpha} + \sigma_{\alpha}^2} \right).
\label{e:vara}
\end{IEEEeqnarray}
The fact that the mean and variance of $a$ are not simply the exponentiation of the 
corresponding $\alpha$ domain values leads to interesting characterisations of stability in
the $x$ domain.

The mode and median of $a$ are also given by simple expressions,
\begin{IEEEeqnarray}{rCl}
\modedist(a) &  = &  \e{\mu_{\alpha} - \sigma_{\alpha}^2} \label{e:modevln}\\*
\noalign{\medskip}
\median(a) &  = & \e{\mu_{\alpha}}.\label{e:medvln}
\end{IEEEeqnarray}
The median condition---and any other quantile value---is transformed via exponentiation making it
a simple matter to characterise properties of the median or quantile value.

\section{Stability conditions}

The following sections derive the conditions under which the median, mean,
and variance of $\fxK{x}$ converge to zero as $K \longrightarrow\infty$.  

\subsection{Mean stability}
\label{s:meanstability}

The condition for the stability of the mean of $x_K$ is a simple consequence of the 
fact that for two independent distributions, the product of the expectations is equal to
the expectation of the product.

\begin{theorem}[Mean stability]
\label{t:lnmeanstable}
\[
\lim_{K\longrightarrow\infty} \mean(x_K) \, = \, 0 
\quad \iff \quad
\mu_a < 1.
\]
\end{theorem}

For lognormal distributions (\ref{e:muv}) shows that the 
mean stability condition can also be stated in terms in the mean
and variance of the $\falpha{\alpha}$ distribution,
\[
\lim_{K\longrightarrow\infty} \mean(x_K) \, = \, 0 
\quad \iff \quad
\mu_{\alpha} + \sigma_{\alpha}^2/2 \, < \,  0.
\]

\subsection{Variance stability}
\label{s:varstability}

Goodman~\cite{goodman:1962a} derives the 
variance of a product of arbitrary distributions which directly leads
to the following variance stability result. 

\begin{theorem}[Variance stability]
\label{t:lnvarstable}
\[
\lim_{K\longrightarrow\infty} \variance(x_K) \, = \, 0 
\quad \iff \quad
\mu_a^2 + \sigma_a^2 < 1.
\]
\end{theorem}

\begin{IEEEproof}
From~\cite{goodman:1962a} we have ,
\begin{IEEEeqnarray*}{rCl}
\variance(x_K)  & = & \variance\left( \prod_{k=0}^{K-1} a_k \right) \\*[0.5em]
		 & =  & \prod_{k=0}^{K-1} \left(\sigma_{a}^2 + \mu_{a}^2\right)
		- \prod_{k=0}^{K-1} \mu_{a}^2,
\end{IEEEeqnarray*}
which in our i.i.d.\ case gives,
\begin{IEEEeqnarray}{rCl}
  \variance(x_K) 
    &  =  &  \left(\sigma_a^2 + \mu_a^2\right)^K - \mu_a^{2K} \label{e:gsum1}  \\*[0.5em]
     & = & \sigma_a^2(\sigma_a^2 + \mu_a^2)^{K-1} \nonumber \\*[0.5em]
     & & +\> \mu_a^{2K} \left( \left(1 + \frac{\sigma_a^2}{\mu_a^2}\right)^{K-1} -1 \right)
     		\IEEEeqnarraynumspace \label{e:gsum2}.
\end{IEEEeqnarray}
If $\sigma_a^2 + \mu_a^2 < 1$ then both terms in~(\ref{e:gsum1}) go to zero as $K\longrightarrow\infty$.
In the case where $\sigma_a^2 + \mu_a^2 =1$ and $\sigma_a^2 > 0$ (\ref{e:gsum1}) also 
shows that the variance of $x_K$ goes to one as $K \longrightarrow\infty$.    If $\sigma_a^2 + \mu_a^2 > 1$
then both terms in (\ref{e:gsum2}) grow without bound as $K \longrightarrow\infty$. 
\hfill \IEEEQEDhere
\end{IEEEproof}

In the lognormal distribution case, substituting~(\ref{e:muv}) and~(\ref{e:vara}) into the
condition of Theorem~\ref{t:lnvarstable} gives
an equivalent condition in terms of the normal $\falpha{\alpha}$ distribution.
\[
\lim_{K\longrightarrow\infty} \variance(x_K) \, = \, 0 
\quad \iff \quad
\mu_{\alpha} + \sigma_{\alpha}^2 < 0.
\]

\subsection{Median stability: lognormal case}

The least restrictive stability condition to be considered is that for
the median of $x_K$.  This result is easy to obtain for a lognormal distribution and
so we do that first.

\begin{theorem}[Median stability; lognormal distribution]
\label{t:lnmedstable}
If $\fa{a}$ is a lognormal distribution,
\[
\lim_{K\longrightarrow\infty} \median(x_K) \, = \, 0 
\quad \iff \quad
\mu_{\alpha}  <  0.
\]
\end{theorem}

\begin{IEEEproof}
As $\fzetaK{x}$ is a normal distribution its median is equal to its mean,
\[
\median(\zeta_k) = \mean(\zeta_K) = K\mu_{\alpha}.
\]
This immediately gives $\lim_{K\longrightarrow\infty} \median(\zeta_K) = -\infty$ if and
only if $\mu_{\alpha} < 0$.   As $\median(x_k) = \e{\median(\zeta_K)}$ the result follows.
\hfill \IEEEQEDhere
\end{IEEEproof}

From~(\ref{e:mualpha}) we can also express the condition of Theorem~\ref{t:lnmedstable}
in terms of the $\fa{a}$ distribution,
\[
\lim_{K\longrightarrow\infty} \median(x_K) \, = \, 0 
\quad \iff \quad
\mu_a^2 - \frac{\sigma_a^2}{\mu_a^2} < 1 .
\]
Note that, depending on the variance $\sigma^2_a$, systems with mean $\mu_a$ greater than one might
still be median stable.  This point will be discussed in greater detail later.

The stability results for lognormal distributions are summarised in Table~\ref{t:lognormalstability}.
All of the conditions can be expressed in terms of the mean and variances of both
the $\fa{a}$ and  the $\falpha{\alpha}$ distributions.

\begin{table}
\centering
\renewcommand{\arraystretch}{1.5}
\begin{tabular}{|l|l|l|}
\hline
{\renewcommand{\arraystretch}{1.1}
\begin{tabular}{l} Stability \\ property \end{tabular} }
         & $\fa{a}$ distribution & $\falpha{\alpha}$ distribution \\
\hline\hline
$\median(x_K)$ &  $\mu_a^2 - \sigma_a^2/\mu_a^2 < 1 $  & $\mu_{\alpha} < 0 $ \\
\hline
$\mean(x_K)$ &  $\mu_a < 1$ & $\mu_{\alpha}  +  \sigma_{\alpha}^2/2  <  0$ \\
\hline
$\variance(x_K)$  & $\mu_a^2  +  \sigma_a^2  <  1$ & $\mu_{\alpha} + \sigma_{\alpha}^2 < 0$  \\
\hline
\end{tabular}
\vspace*{6pt}
\caption{\label{t:lognormalstability}Stochastic feedback gain stability conditions
for lognormal distributions}
\end{table}%

\subsection{Median stability: general distributions}
\label{s:medianstability}

We now consider median stability in the case where the $\fa{a}$ distributions
are other than lognormal.  The mean and variance relationships between
the $\fa{a}$ and $\falpha{\alpha}$ distributions given in Equations~(\ref{e:mualpha}) to~(\ref{e:vara})
no longer hold.  Unfortunately this is also true for all values of $K$ and also in 
the limit as $K\longrightarrow\infty$.

The situation is more complex for more general $\fa{a}$ distributions
as $\falpha{\alpha}$ is not normal.   In the context of the central limit theorem, it is perhaps
surprising that although the distribution $\fzetaK{\zeta}$ is the $K$-fold convolution of
the $\falpha{\alpha}$ distributions,  the median of $\fzetaK{\zeta}$ does not necessarily converge
to the mean of $\fzetaK{\zeta}$.  The difference can be quantified.

\begin{lemma}
\label{l:medianlimit}
Assume that $\falpha{\alpha}$ is a non-lattice distribution with bounded third moment.  Then
\[
\lim_{K\longrightarrow\infty} \: \median(\zeta_K) \, - \,  \dexpec{ \zeta_K }
	\: = \: - \frac{ \dexpec{ (\alpha - \mu_{\alpha})^3 } } { 6 \sigma_{\alpha}^2}.
\]
\end{lemma}	
A lattice distribution is one where
there exist parameters $b\in\R$ and $h>0$ such that $\Prob\{\alpha \in b + h\Z\} = 1.$

\noindent
\begin{IEEEproof}
\begin{IEEEeqnarray*}{rCl}
\median(\zeta_K) - \dexpec{ \zeta_K }  
           &  = & \median\left(\sum_{k=0}^{K-1} \alpha \right) - K\mu_{\alpha}  \\
             & = &  \sigma_{\alpha} \median\left( \sum_{k=0}^{K-1} \frac{\alpha - \mu_{\alpha}}{\sigma_{\alpha}} \right).
\end{IEEEeqnarray*}
Define a new stochastic variable $y$  with probability density function $f_y(y)$ by,
\[
y  := h(\alpha)  =  \frac{\alpha - \mu_{\alpha}}{\sigma_{\alpha}},  
\]
and note that $\dexpec{y} = \mu_y = 0$ and $\dexpec{y^2} = \sigma_y^2 = 1$.   Define $\tau$ as the third moment of $f_y(y)$,
\[
\tau  :=   \dexpec{ y^3 }, 
\]
and by assumption $\tau < \infty$.  The following result is from Hall~\cite{hall:1980a}
provides the
key step.   If  $f_y(y)$ is a non-lattice distribution with $\mu_y = 0$, $\sigma_y =1$ and
$\dexpec{ y^3 } = \tau < \infty$,
then,
\[
\lim_{K\longrightarrow\infty}  \median\left( \sum_{k=0}^{K-1} y \right) \: = \: \frac{-\tau}{6}.
\]
As
\begin{IEEEeqnarray}{rCl}
\IEEEeqnarraymulticol{3}{l}{
\lim_{K\longrightarrow\infty} \: \median(\zeta_K) \, - \,  \dexpec{ \zeta_K } }  \nonumber \\*
	\qquad &  = & \sigma_{\alpha} \median\left( \sum_{k=0}^{K-1}
			 \frac{\alpha - \mu_{\alpha}}{\sigma_{\alpha}} \right) \nonumber \\
		&  = & \sigma_{\alpha} \median\left( \sum_{k=0}^{K-1}y \right) \: = \: \frac{-\sigma_{\alpha}\,\tau}{6}, \label{e:rawmedlim}
\end{IEEEeqnarray}
it only remains to determine the value of $\tau$.   As $h^{-1}(y) = \sigma_{\alpha} y + \mu_{\alpha}$,
\begin{IEEEeqnarray*}{rCl}
\tau &  = & \int_{-\infty}^{\infty} y^3 f_y(y) \dd y  \\
      &  = &  \int_{-\infty}^{\infty} y^3  \falpha{h^{-1}(y)} \left| \frac{d h^{-1}(y)}{dy} \right|\, \dd y\\
    &  = & \int_{-\infty}^{\infty} y^3  \falpha{\sigma_{\alpha} y + \mu_{\alpha}} | \sigma_{\alpha}|\, \dd y \\
   &   = &   \int_{-\infty}^{\infty} \frac{ (\alpha - \mu_{\alpha})^3}{\sigma_{\alpha}^3} 
    	\falpha{\alpha} |\sigma_{\alpha}| \, \frac{\dd \alpha}{|\sigma_{\alpha}|} \\
    &  = & \frac{1}{\sigma_{\alpha}^3} \int_{-\infty}^{\infty} (\alpha - \mu_{\alpha})^3 \falpha{\alpha} \dd \alpha \\
    &  = & \frac{ \dexpec{ (\alpha - \mu_{\alpha})^3 } }{\sigma_{\alpha}^3}.
\end{IEEEeqnarray*}
Substituting the above into~(\ref{e:rawmedlim}) gives the desired result. \hfill \IEEEQEDhere
\end{IEEEproof}

The key point in determining the median stability is that the limit in Lemma~\ref{l:medianlimit}
is independent of $K$.

\begin{theorem}[Median stability)]
\label{t:medstab}
Assume that $\falpha{\alpha}$ is a nonlattice distribution with bounded third moment.  Then,
\[
\lim_{K\longrightarrow\infty} \median(x_K) \; = \;  0
\quad \iff \quad
	\mu_{\alpha} \; < \; 0.
\]	
\end{theorem}

\begin{IEEEproof}
If $\mu_{\alpha} = 0$ then $\dexpec{ \zeta_K } \; = \; K\mu_{\alpha} \; = \; 0$,
and from Lemma~\ref{l:medianlimit}, $\lim_{K\longrightarrow\infty} \median(\zeta_K)$ is a finite constant.
Therefore,
\[
\lim_{K\longrightarrow\infty}  \median(x_K)  \; = \;  \lim_{K\longrightarrow\infty}  \e{\median(\zeta_K)} \; \neq \; 0.
\]
By Lemma~\ref{l:medianlimit}, for every $\epsilon > 0$, there exists an integer $\bar{K}$
such that for all $K > \bar{K}$,
\[
\left| \median(\zeta_K) - K\mu_{\alpha}   + \frac{ \dexpec{ (\alpha - \mu_{\alpha})^3 } } { 6 \sigma_{\alpha}^2} \right| \: < \: \epsilon.
\]
This implies that
\[
| \median(\zeta_K) - K\mu_{\alpha} | \: < \:  \left| \frac{ \dexpec{ (\alpha - \mu_{\alpha})^3 } } { 6 \sigma_{\alpha}^2} \right| \, + \, \epsilon.
\]
If we now assume that $\mu_{\alpha} < 0$ then $K\mu_{\alpha} < 0$ and
\[
\median(\zeta_K) \: < \:  K\mu_{\alpha} +  \left| \frac{ \dexpec{ (\alpha - \mu_{\alpha})^3 } } 
	{ 6\, \sigma_{\alpha}^2} \right| \, + \, \epsilon.
\]
The righthand side clearly goes to $-\infty$ as $K\longrightarrow\infty$.  An analogous argument
for $\mu_{\alpha} > 0$ gives a lower bound on $\median(\zeta_K)$ that goes to $\infty$ as $K\longrightarrow\infty$.
Exponentiating $\median(\zeta_K)$ gives the required result.
\hfill \IEEEQEDhere
\end{IEEEproof}

This result also follows from Cantelli's inequality (see Lemma~\ref{l:cantelliBound}) without the
requirement of a bounded third moment.   However the method of proof above illustrates the manner
in which a sum of non-normal distributions does not converge to a normal distribution.  It also gives the
following interesting boundary condition.

\begin{corollary}[Median limit: zero mean log distribution]
If $\falpha{\alpha}$ is a non-lattice distribution with $\mu_{\alpha} = 0$ and
$ \left| \dexpec{ (\alpha - \mu_{\alpha})^3 } \right| < \infty$
then,
\[
\lim_{K\longrightarrow\infty} \median(x_K)  
	\: = \: \e{\displaystyle  - \frac{ \dexpec{ (\alpha - \mu_{\alpha})^3 } } { 6\, \sigma_{\alpha}^2} }.
\]	
\end{corollary}

Ethier~\cite{ethier:2004a} also uses the result of Hall to prove a similar result on the median of
a gambler's fortune.  The results in~\cite{ethier:2004a} suggest that the assumption of a non-lattice
distribution in Lemma~\ref{l:medianlimit} may be able to be removed.  Such distributions are important
for gambling applications but may be of less interest in many control applications.

The stability conditions for more general distributions (non-lattice and with bounded third
moment) are summarised in Table~\ref{t:stability}.  

\begin{table}
\centering
\renewcommand{\arraystretch}{1.5}
\begin{tabular}{|l|l|l|}
\hline
{\renewcommand{\arraystretch}{1.1}
\begin{tabular}{l} Stability \\ property \end{tabular} }
         & $\fa{a}$ distribution & $\falpha{\alpha}$ distribution \\
\hline\hline
$\median(x_K)$ &  --- & $\mu_{\alpha} < 0 $ \\
\hline
$\mean(x_K)$ &  $\mu_a < 1$ & --- \\
\hline
$\variance(x_K)$  & $\mu_a^2 + \sigma_a^2 < 1$  & --- \\
\hline
\end{tabular}
\vspace*{6pt}
\caption{\label{t:stability}Stochastic feedback gain stability conditions for
more general distributions}
\end{table}%

The median stability conditions given here for a stochastic gain $a_k \sim \fa{a}$ have
a similar form to the stability conditions for a time-varying gain.  
Section~\ref{sb:periodic}  gives more detail on this perspective.

\subsection{Stabilisation by time-varying gains and the geometric mean}
\label{sb:periodic}

Some of the initially unintuitive phenomena observed for stochastic feedback may be better 
understood by considering systems with certain types of deterministic, but time-varying feedback gains. 
For the case of a scalar state, a complete analysis is easy to do.  For a more complete analysis of the 
periodic multivariable case see for example~\cite{bittanti:2009a}.
Consider the single-state discrete-time system and its solution 
\be
x_{k+1} ~=~  a_k ~x_k  
\quad
\Longrightarrow 
\quad
x_K ~=~ \left( \prod_{k=0}^{K-1}   a_k  \right) x_0.		
\label{disceq.eq}
\ee

 If $\bu$ is a periodic signal with period $N$, then the growth of $x$ can be characterised by observing
the behaviour every $N$ time steps. Define the ``sub-sampled state''  
\[
\hat{x}_k ~:=~ x_{kN}. 
\]
Note that $x$ decays iff $\hat{x}$ decays since  the growth of $x$ in between the subsamples is bounded. 
The recursion for $\hat{x}$ is time invariant 
\[
\hat{x}_{k+1}  ~=~ 
		x_{(k+1) N}  ~=~  \left( \prod_{k=0}^{N-1} \bu(k)   \right)  x_{kN} 
		~=:~ \hat{a}  ~\hat{x}_{k},
\]
where 
$ \hat{a} := \prod_{k=0}^{N-1}  a_k $  is the so-called  ``monodromy gain''. 
Thus the sequence $\hat{x}$ decays iff 
\be
|\hat{a}| <1 
\quad \Leftrightarrow \quad
|\hat{a}|^{\frac{1}{N}} < 1
\quad \Leftrightarrow \quad
 \left( \prod_{k=0}^{N-1} |a_k|   \right)^{\frac{1}{N}}  < 1.  \label{geo_mean.eq}
\ee
		 
\begin{figure*}         
\begin{minipage}[t]{0.7\linewidth}
\strut\vspace*{-\baselineskip}\newline\includegraphics{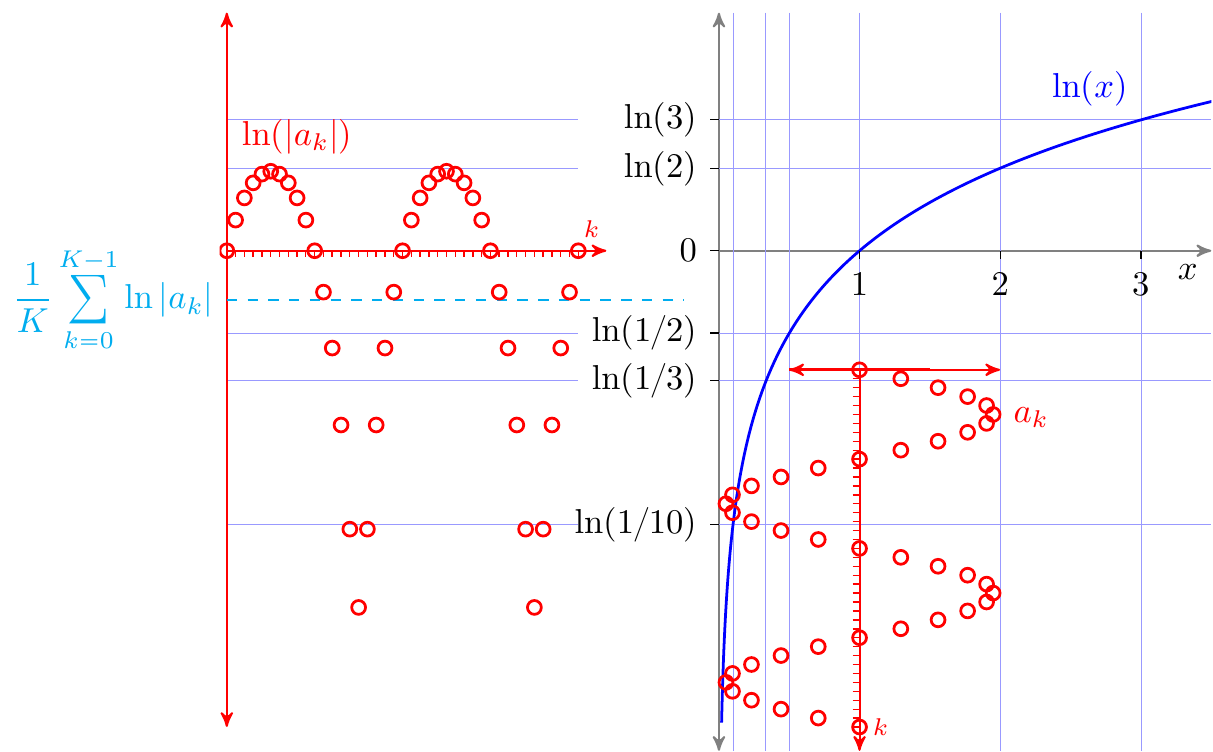} 
\end{minipage}
\hfill
\begin{minipage}[t]{0.275\linewidth}
\caption{\label{f:LogTrans} An illustration of the mapping from $a_k$ to $\ln|a_k|$ showing that when $a$ 
is symmetrically distributed around $1$, $\ln|a|$ is distributed more heavily towards negative numbers
due to the distortion by the $\ln$ mapping. The 
cyan dashed line on the graph of $\ln|a|$ 
indicates the arithmetic mean of that signal (this is the logarithm of the geometric mean of $a$)	
showing how it is negative while the $\ln$ of the mean of $a$  is zero.}
\end{minipage}
\end{figure*}

The  last quantity in~\req{geo_mean} is the {\em geometric mean} of the absolute value
of the  signal $\bu$, which is 
the right quantity that characterises stability in this system. The geometric mean can also 
be expressed using the arithmetic mean of the 
logarithm,
\begin{IEEEeqnarray*}{rCl}
 \left( \prod_{k=0}^{N-1} |a_k|   \right)^{\frac{1}{N}}  < 1  
   \quad & \Leftrightarrow & \quad 
\ln\left( \prod_{k=0}^{N-1} | a_k | \right)^{\frac{1}{N}} \!\!\!\! < \,0  \\
	 & \Leftrightarrow & \quad
	\frac{1}{N} \sum_{k=0}^{N-1} \ln\big| a_k \big| ~<~0 .
\end{IEEEeqnarray*}
Thus the system is asymptotically stable iff the {\em arithmetic mean of $\left\{ \ln | \bu_k | \right\}$ is negative}. 
Note how this is analogous to the condition $E\{ \ln|a| \}<0$ when $a$ is a stochastic process. 
The relation between the geometric mean and the arithmetic mean through the logarithm function is illustrated 
in Figure~\ref{f:LogTrans}.
The figure illustrates a periodic gain $a$ that is symmetrically 
distributed around 1. The $\ln(a_k)$ mapping tends to boost values of $\{a_k\}$ that are less than 1 more 
heavily towards large negative numbers while tempering the values of $\{a_k\}$ that are larger than 1
by mapping them to smaller positive numbers. The result is that even though $a$ maybe symmetrically 
distributed around 1, the product $\prod_{k=0}^{N-1} \left| a_k \right| $ will be strictly smaller than 1. 


Now let's examine~\req{disceq} in the case where the sequence $\{a_k\}$ is a general time-varying gain. 
The asymptotic behavior of the solution is completely determined by the  limit of the product  
of the gains $\{a_k\}$, which can be studied as a series limit 
by taking the 
logarithm,
\[
\ln \left| \prod_{k=0}^{K-1} a_k \right| ~=~ \sum_{k=0}^{K-1} \ln\big| a_k \big| .
\]
Now  explore the limit 
\begin{IEEEeqnarray*}{rCl}
\lim_{K\rightarrow\infty}
\sum_{k=0}^{K-1} \ln\big| a_k \big| 
	 & =  & 
\lim_{K\rightarrow\infty} ~
K \left( \frac{1}{K} \sum_{k=0}^{K-1} \ln\big| a_k \big| \right)  \\
 	& =  & 
K \left( 
\lim_{K\rightarrow\infty} ~
\frac{1}{K} \sum_{k=0}^{K-1} \ln\big| a_k \big| \right) \\
	   & =:  & 
K~{\cal E}\left[ \ln|\bu| \right] , 
\label{asav.eq}		  
\end{IEEEeqnarray*}
where the last limit is expressed in terms of the 	
{\em asymptotic average}, which for any signal $u$ is defined  by 
\[
{\cal E} \left[ u \right]  := \lim_{K\rightarrow\infty} \frac{1}{K}  \sum_{k=0}^{K-1} u_k.
\] 
This asymptotic average 
can be thought of as a ``deterministic expectation'' of $u$, which is equivalently the time
average of a realization of a stochastic process. 

We can thus conclude that if the asymptotic average of $\ln|\bu|$ exists and  is negative, then 
the state will asymptotically converge to zero, i.e. 
\be
\dexpec{ \ln\left|\bu \right|} ~<~ 0 
\quad \Rightarrow \quad
\lim_{K\rightarrow\infty} x_K ~=~ 0. 
\label{logcond.eq}
\ee
We can actually conclude something 
slightly stronger,
\be
\dexpec{\ln|\bu| }  ~=:~ \ln(\gamma) ~<~ 0 
\quad \Rightarrow \quad
\left| x_K  \right|  ~\leq~ \alpha ~\gamma^K , 
\label{geoconv.eq}		  
\ee
i.e.\ the convergence is geometric with decay rate  $\gamma<1$.  Finally we note that condition~\req{geoconv}
is only necessary for exponential convergence. Slower convergence can still occur even when this
condition does not hold. For convergence we simply need the sequence on the right hand side
of~\req{asav} to go to $-\infty$. This can occur even when the asymptotic average is converging to 
zero (from below), 
as long as it converges at a rate slower than $1/K$. More precisely we can state 
\begin{multline*}
\lim_{K\rightarrow \infty} x_K = 0  \\
 \Leftrightarrow  \quad 
\lim_{K\rightarrow\infty} ~K
\left( 
\frac{1}{K} \sum_{k=0}^{K-1} \ln\big| \bu_k \big| \right) 
 = -\infty.
\end{multline*}

\subsection{Stability regions}

\begin{figure}
\includegraphics[width=\linewidth]{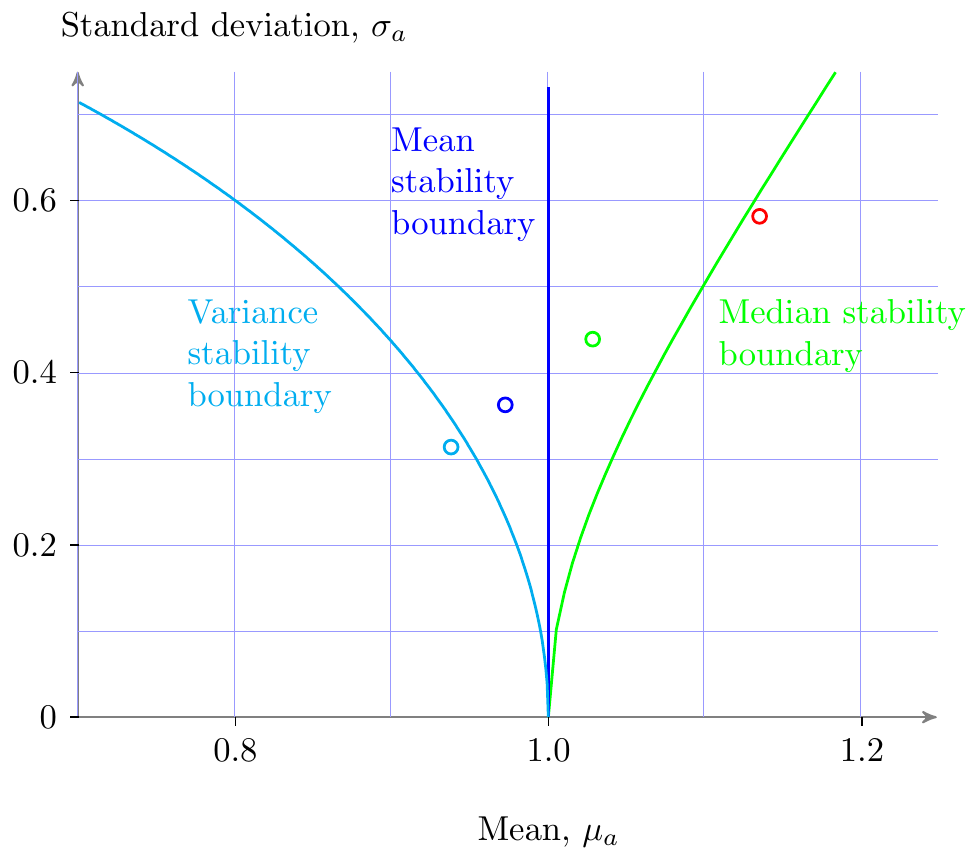}
\caption{\label{f:vdombdy} Stability regions in the $a$ domain.  The region where the condition
is satisfied is to the left of the correspondingly coloured boundary.  The mean and variance 
stability regions are applicable to general distributions.  The median stability region shown here
applies to lognormal $\fa{a}$ distributions.  Other distributions will have slightly different median
stability regions, but identical mean and variance stability regions.  The four circles indicate 
the values of $\mu_{\alpha}$ and $\sigma_{\alpha}$  of the distributions shown in
Figure~\protect\ref{f:vlndistcases}.}
\end{figure}

The variance, mean, and median stability regions for the $\fa{a}$ distribution are
shown in Figure~\ref{f:vdombdy}.   The most interesting observation is that there 
exists a region in which $\median(x_K$) is stable and the $\mean(x_K)$ is unstable.
We will subsequently show (in Section~\ref{s:concentration})
that in this case the sample paths $x_K$ go to zero 
almost surely but the mean of $x_K$ goes to $\infty$.   This analysis can also be applied
to other feedback loops and Section~\ref{sb:stabilisation} derives the stability regions for a first order 
system with unknown gain and pole position.

\begin{figure}
\includegraphics[width=\linewidth]{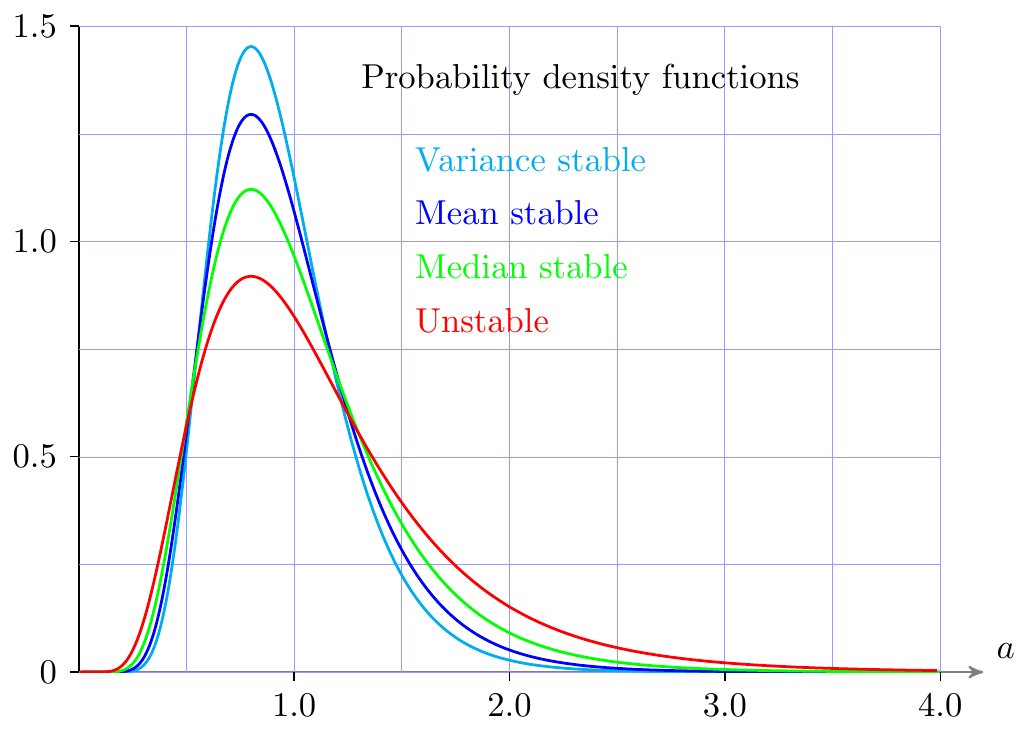}
\caption{\label{f:vlndistcases} Probability density functions of $a$ for four
stability cases: variance stable, mean stable, median stable, and unstable.  All
four cases have the same mode;  their means and standard deviations are shown in
Figure~\protect\ref{f:vdombdy}. }
\end{figure}

Figure~\ref{f:vlndistcases} gives the probability distribution functions for four
stability cases: unstable, median stable, mean stable, and variance stable.
In all cases the mode of the distribution is less than one.  The remarkable feature of these
distributions is that they are not particularly different and yet give very different stability
characteristics in the evolution of the state.

\begin{figure}
\includegraphics[width=\columnwidth]{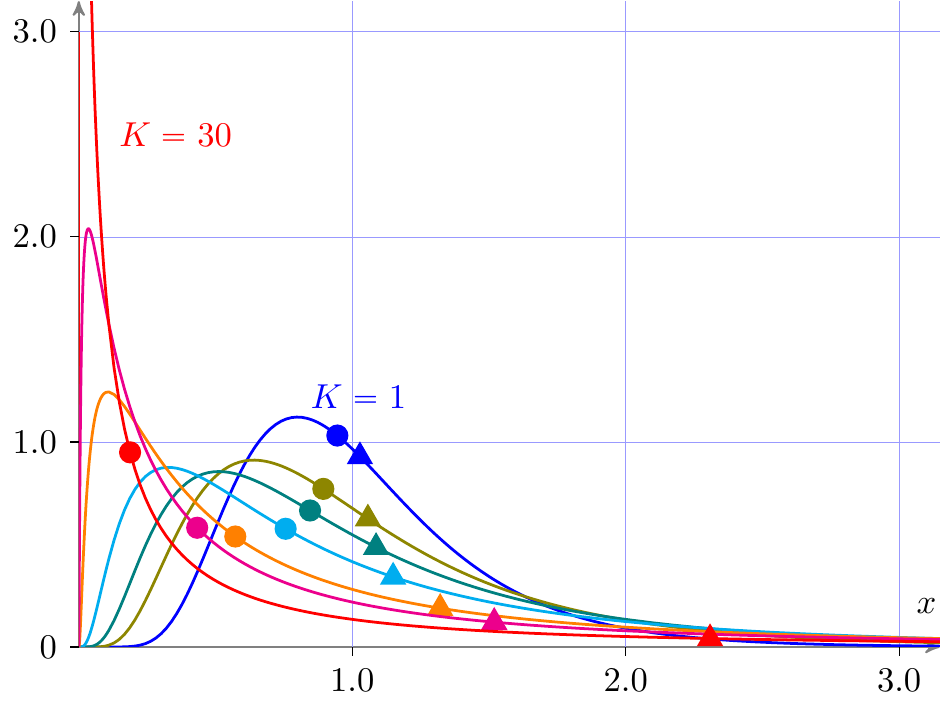}
\caption{\label{f:xKdistevolve} Evolution of the probability density function of $x_K$ 
for $K = 1, 2, 3, 5, 10, 15$, and $30$.  The mean values of each $x_K$ distribution are indicated by triangles
and the median values are indicated by solid dots. As $K\longrightarrow \infty$ the mean of $x_K$ increases 
and the median decreases.}
\end{figure}

The most intriguing case is that where the median of $x_K$ is stable, but the mean is unstable. 
Figure~\ref{f:xKdistevolve} shows the evolution of the log-normal probability density function of $x_K$ for 
a range of values of $K$.  The evolution of the median towards zero, and the mean towards infinity, 
are clear in the distributions.

\begin{figure*}
\begin{minipage}[t]{0.65\textwidth}
\vspace*{0pt}
\strut\vspace*{-\baselineskip}\newline\includegraphics{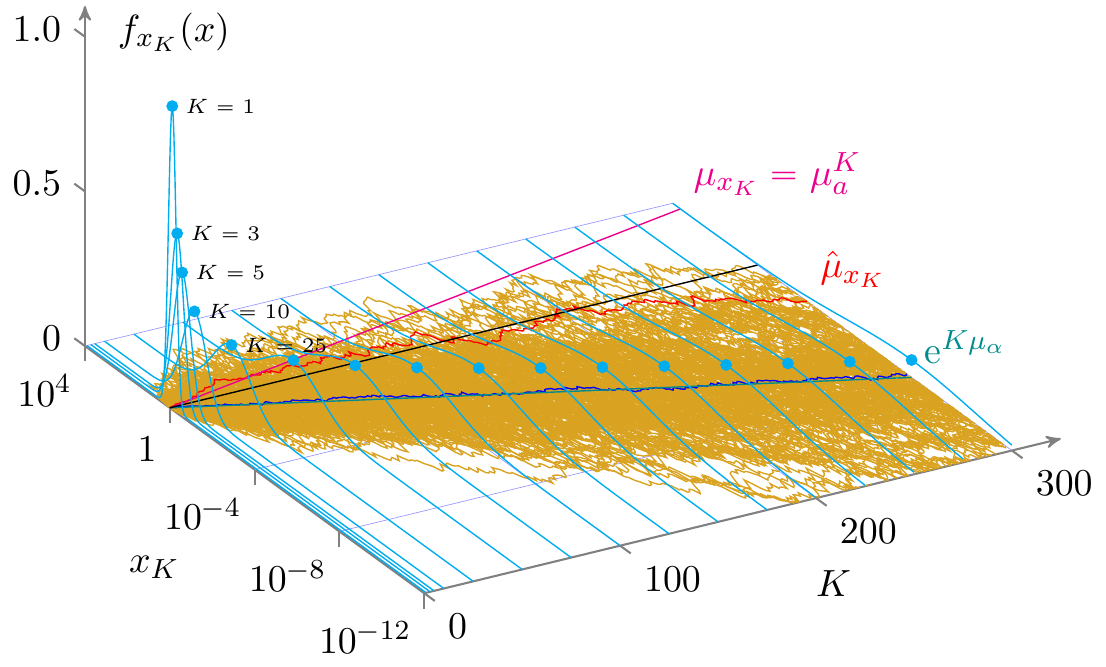}
\end{minipage}
\hfill
\begin{minipage}[t]{0.325\textwidth}
\vspace*{0pt}
\caption{\label{f:lognormcase3} Simulation of the median stable/mean unstable 
case.  The vertical axis shows the evolution of the probability density function, $f_{x_K}(x)$ for a range
of values of $K$.  The horizontal plane shows 200 simulation sample paths of $x_K$ as a function
of $K$.   Also shown are the sample median (blue line) and
sample mean (red line: $\hat{\mu}_{x_K}$), along with the
theoretical median (cyan line: $\e{K\mu_{\alpha}}$) and theoretical mean (magenta line: $\mu_{x_K}$).
The theoretical values are derived by mapping the $\alpha$-space distribution in~(\protect\ref{e:zetadist})
through the equations~(\protect\ref{e:medvln}) and~(\protect\ref{e:modevln}).}
\end{minipage}
\end{figure*}

We illustrate the median stable/mean unstable 
case by simulating 200 sample
paths.   The $\fa{a}$ distribution is the lognormal distribution with probability density function shown as Case~3 of
Figure~\ref{f:vlndistcases} ($\mu_a = 1.0283$, $\sigma_a = 0.4389$).
Figure~\ref{f:lognormcase3} illustrates the sample paths and the evolution of the distribution,
$\falpha{\alpha}$.   As $K$ increases
the median stability condition ensures that the sample paths go to zero almost
surely.   However $\mu_a > 1$ and so the $\mean(x_K)$ is unstable and goes
to $\infty$.   For $K$ very large this results in 
an $x_K$ distribution with a very high peak close to $x_K = 0$, but still having enough weight
in the positive tail that the mean of $x_{K}$ is very large (and growing with $K$).   The sample 
estimate of $\mean(x_K)$ (denoted by $\hat{\mu}_{x_K}$) drops below the theoretical mean, $\mu_{x_K}$
as increasingly fewer sample paths are near or above the mean.  This phenomenon is 
investigated in more detail in Section~\ref{s:concentration}.


\subsection{Stochastic gain stabilisation}
\label{sb:stabilisation}

\begin{figure}
\centering
\includegraphics{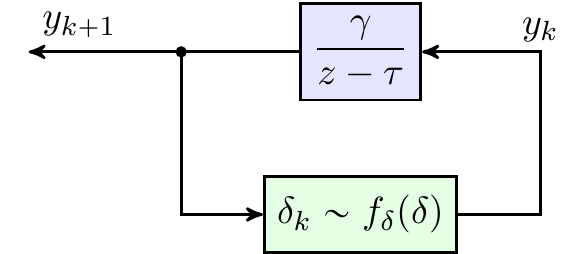}
\caption{\label{f:stochasticfb} Stochastic stabilisation problem.  A first order plant is connected in
feedback with a stochastic gain $\delta_k \sim \fdelta{\delta}$. 
}
\end{figure}

The stability results for stochastic feedback can easily be applied to the slightly more problem of
stabilising a general first order system via stochastic feedback.  Figure~\ref{f:stochasticfb} illustrates
the configuration for this problem.     This is a simple case of a more general stochastic stabilisation problem, 
referred to as stabilisation by noise.  This problem has been studied in stochastic vibration control
context (see the review~\cite{Roberts:1986a}).  In vibration control the assumption of
an oscillatory nominal response is usually exploited.  The more general case is studied in~\cite{Arnold:1983a}
and is based on earlier work in~\cite{oseledec:1968a}.  This work focuses the continuous-time equivalent
to the mean stability case considered in this paper.   The application example given here has
also been studied in~\cite{milisavljevic:1997a}, where a result which is essentially equivalent to
the median stability boundary below is presented.  

The plant is given and has the transfer function $G(z)$,
\[
G(z) \: = \: \frac{\gamma}{z - \tau}.
\]
Denote the plant output by $y_k$.  The closed loop dynamics of the feedback system 
illustrated in Figure~\ref{f:stochasticfb} are given by,
\[
y_{k+1} \: = \: \left( \tau + \gamma \delta_k \right) \, y_k, 
\]
where $\delta_k \sim \fdelta{\delta}$ is the stochastic feedback drawn from a known distribution
at each time instant.

Now define $x_k$ = $|y_k|$ and note that,
\[
x_{k+1} \: = \:  | \tau + \gamma \delta_k | \, x_{k}.
\]
As $x_k \geq 0$ for all $k$, the results summarised in Table~\ref{t:stability} are directly applicable by defining
\[
a_k \: = \: | \tau + \gamma \delta_k |.
\]
The mean of the $\fa{a}$ distribution is,
\[
\mu_a \: = \: | \tau + \gamma \mu_{\delta} |.
\]
The variance may be more difficult to evaluate precisely but can be easily estimated numerically.  
If the distribution $\fdelta{\delta}$ were such that $\delta_k > 0$ then,
\[
\sigma^2_{a} \; = \; \gamma \sigma^2_{\delta}.
\]
However the absolute value in the definition of $a$ complicates this somewhat, particularly in the 
case of interest where $\tau \neq 0$.   

\begin{figure}
\includegraphics[width=\columnwidth]{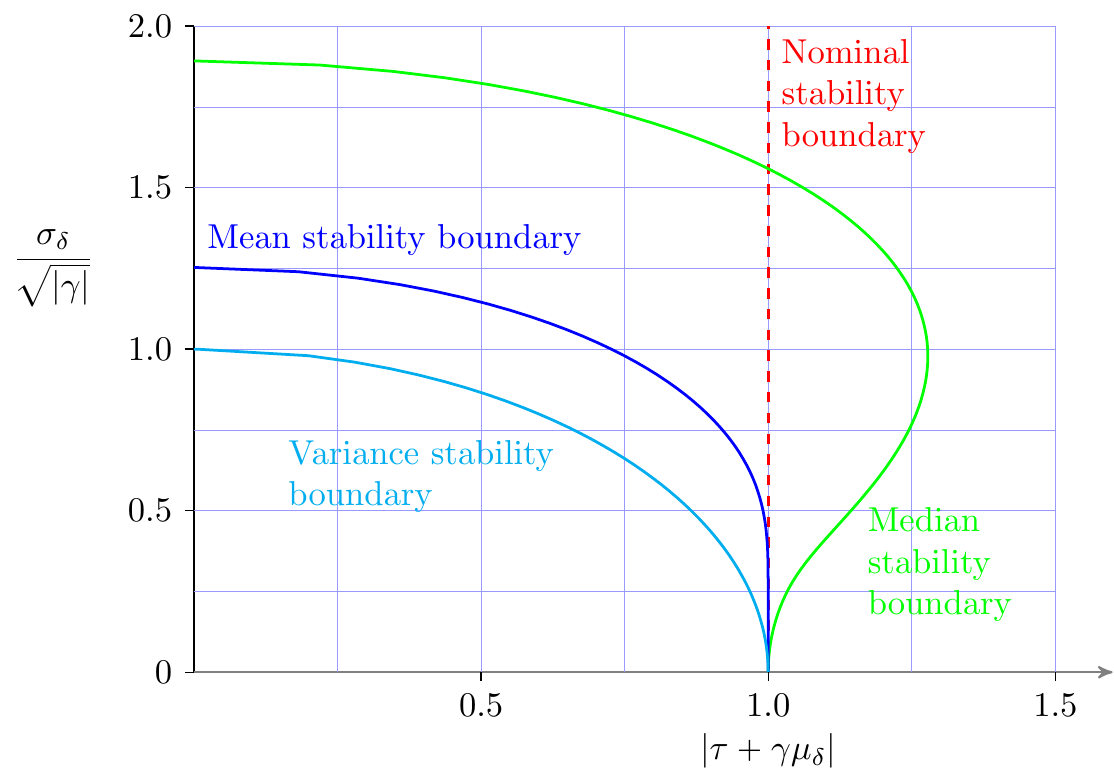}
\caption{\label{f:stochStabBdy}  Stability boundaries for the state-magnitude evolution for the 
plant, $G(z) = \gamma/(z - \tau)$, in feedback with a stochastic gain,
$\delta_k \sim \rvN(\mu_{\delta},\sigma^2_{\delta})$.    Unstable open-loop plants may be median
stabilised by a stochastic gain with the appropriate mean and variance.
}
\end{figure}

As expected the conditions for median, mean, and variance stability differ and for a given distribution,
$\fdelta{\delta}$, a stability boundary diagram, analogous to that in Figure~\ref{f:vdombdy}, can be drawn.  
Figure~\ref{f:stochStabBdy} illustrates the stability regions for the case where $\delta_k$ is drawn from 
a normal distribution, $\delta_k \sim \rvN(\mu_{\delta},\sigma^2_{\delta})$.

An $\fdelta{\delta}$ distribution with a non-zero mean can be viewed as a constant feedback gain
of $\mu_{\delta}$ in parallel with a zero-mean stochastic gain.  The static feedback effect of $\mu_{\delta}$
is accounted for in the stability boundary figure by plotting the nominal case as $| \tau + \gamma \mu_{\delta}|$.
Similarly, the standard deviation of the stochastic feedback is scaled by $1/\sqrt{|\gamma|}$ to normalise for
the gain scaling effect of $\gamma$.

The condition for the  nominal stability of the plant is that $|\tau + \gamma \mu_{\delta} | < 1$.   The median
stability boundary shows that for a range of variance, the median of $|y_k|$ is stable.   Note however, the if the
nominal plant is not stable, then neither the mean, nor the variance, of $|y_k|$ can be stabilised by stochastic
feedback.  It is also interesting to note that for any given nominal stability margin there are increasingly large 
values of the variance of the stochastic feedback that will destabilise the variance, mean, and median, in that order.

Another observation is that the stability boundaries involve the absolute values of functions of the
plant parameters $\tau$ and $\gamma$.   This has an interesting robustness interpretation and implies that 
in the $\mu_{\delta} = 0$ case the plant can be median stabilised for a range of $\tau$ and $\gamma$ irrespective of
their signs.  For example for a plant with $\tau$ in the range $-1.05 \leq \tau \leq 1.05$ there exists a zero-mean
normally distributed stochastic feedback of a certain variance which will median stabilise the plant.

This exceptional robustness should not be interpreted as an indication that the stochastic controller is practical.
The mean and variance of the realisations of the trajectories are still growing without bound and the random
excursions could be extremely large.   The stochasticity in the feedback loop leads to distributions of $y_k$
which are heavy-tailed.  The potential value of these results is in avoiding the case where stochasticity in a 
feedback loop inadvertently  leads to destabilisation.

\section{Cumulative Distributions and Concentration Results}
\label{s:concentration}

The above observation, that in the median stable case,
the mass of the distribution falls below the mean,
is examined in more detail.   More specifically we would like to calculate,
or at least provide an upper bound for, the probability that $x_K$ exceeds
a certain value.    Denote the complementary cumulative distribution function
by,
\begin{IEEEeqnarray*}{rCl}
\cFxK{\xbnd}   &  = &  \Prob\{ x_K > \xbnd\}  \\*[0.5em]
                       &  = &  \Prob\left\{ \prod_{k=1}^K a_k > \xbnd \right\} ,
\end{IEEEeqnarray*}
where we haved assumed that $x_0 = 1$.  Furthermore we are interested
in the properties of $\cFxK{\xbnd}$ as $K\longrightarrow \infty$ as this gives us
information about the mass of the distribution of $x_K$ as $K$ increases.

Results of this nature are referred to as concentration inequalities in the statistics literature
and have a long history.   For a much more extensive treatment of concentration inequalities
in stochastic processes similar to the ones considered here see~\cite{bercu:2015a}.

We assume that $\mu_{\alpha} < 0$ (median stable case) and observe that
two choices of $\xbnd$ are of potential interest:
\begin{itemize}
\item[1.]$\xbnd = 1$.
 This gives the probability that $x_K > x_0$.  This addresses the question of the
 probability that a realisation of the $x_k$ trajectory decays over the interval $[0,K]$.\\*[-0.5em]
\item[2.]$\xbnd = \mean(x_K) = \mu_a^K$.
This provides insight into our ability (or lack thereof) to estimate the mean
of $x_K$ from a finite number of sample path realisations.
\end{itemize}
For simplicity this paper will focus on the first case.  The results are easily extended
to the second at the expense of more complex formulae in some cases.

The analysis is of course easier in the $\alpha$-space and so we consider,
\[
\Prob\{ x_K > 1\} \; = \; \Prob\{ \zeta_K > 0 \}.
\]
The objective is to provide bounds on this probability as a function of $K$.

\subsection{Log-normal distribution case}

We first consider the lognormal $\fa{a}$ case as exact formulae are easily derived.
In this case,
\[
\falphaK{\alpha} \: = \: \fN{\alpha,K\mu_{\alpha},K\sigma^2_{\alpha}},
\]
\[
\FalphaK{\alpha} \: = \: \int_{-\infty}^{\alpha} \fN{y,K\mu_{\alpha}, K\sigma^2_{\alpha}} \dd y,
\]
and 
\[
\cFalpha{\alpha} \: = \: \frac{1}{2} 
	\left( 1 - \erf{ \frac{\alpha - K\mu_{\alpha}}{\sqrt{2 K \sigma^2_{\alpha}} } } \right),
\]
where $\erf{x}$ is the error function.  The tail probability is then,
\begin{equation}
\label{e:lognormtail}
\Prob\{ x_K > 1\} \; = \; \frac{1}{2} 
	\left( 1 - \erf{ \frac{-\sqrt{K}\mu_{\alpha}}{\sqrt{2 \sigma^2_{\alpha}} } } \right). 
\end{equation}
In the median stable case $\mu_{\alpha} < 0$ and so the argument of the error function is positive.	

\begin{figure}
\includegraphics[width=\columnwidth]{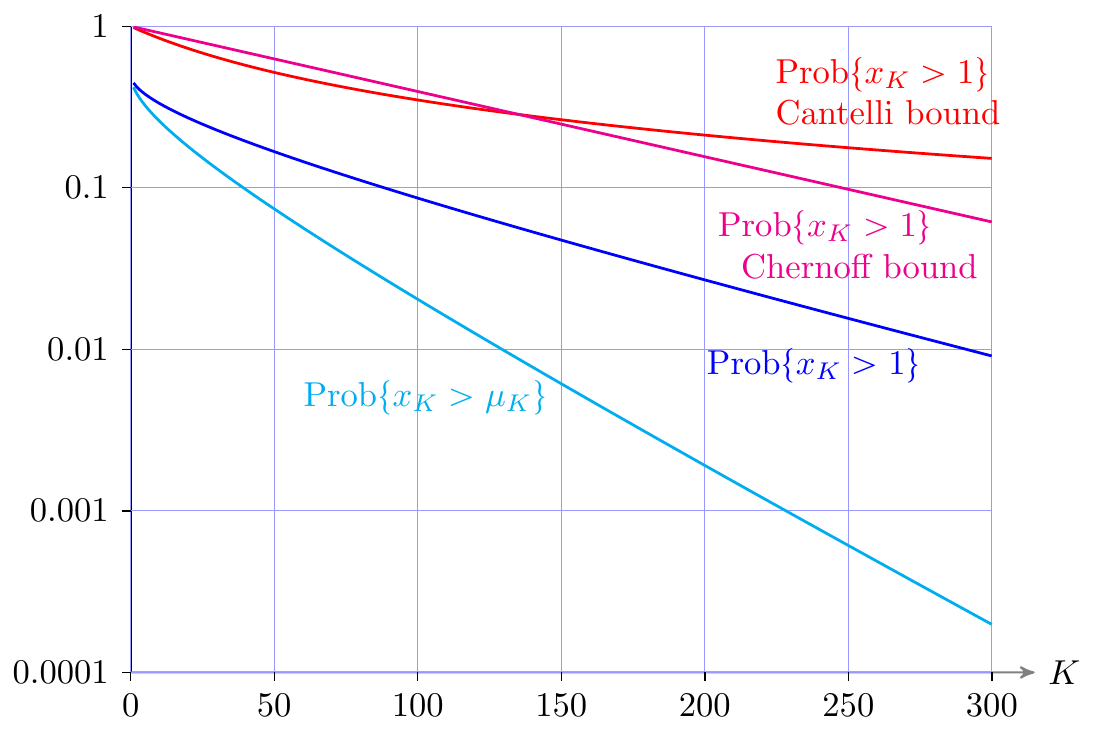}
\caption{\label{f:concbnds}  Concentration inequalities and tail distributions for $x_K$.
The probability that $x_K > 1$ is calculated for the lognormal distribution case studied
in Figure~\ref{f:lognormcase3} (blue line).  Also shown is the probability that $x_K > \mu_{x_K}$
(cyan line).   For comparison several concentration inequalities are also illustrated: the Cantelli inequality
(red), and Chernoff inequality (magenta) for the same distribution.
For the simulation shown in
Figure~\protect\ref{f:lognormcase3} the probability that a sample trajectory will exceed
the mean ($\mu_{x_K} = \mu_a^K$), at the end of the interval ($K=300$) is 0.0002.
}
\end{figure}

Figure~\ref{f:concbnds} illustrates the application of the bound in~(\ref{e:lognormtail}) to the example 
simulated in Figure~\ref{f:lognormcase3}.    When the distribution is known,
$
\Prob\{ x_K \, > \, 1 \}
$
can be calculated numerically and this probability is shown as a function of $K$ for the 
$\fa{a} \sim \rvLN$ case.  Also shown is
\[
\Prob\{ x_K \, > \,  \mu_{x_K}  \} \: = \: \Prob\{ x_K \, > \,  \mu_{a}^K \},
\]
and the exponential decrease of this probability illustrates why the sample-based
estimate of $\mu_{x_K}$ rapidly deteriorates with
increasing $K$. 

At first glance the exponential decay in Figure~\ref{f:concbnds} might seem
counter-intuitive as the complementary cumulative distribution of a standard
normal distribution satisifies,
\[
\cFalpha{\alpha} \: <  \: \frac{1}{\sqrt{2\pi} \alpha} \e{\frac{-\alpha^2}{2}},
\]
for all $\alpha > 0$, which appears to be a significantly faster decay.
However (\ref{e:lognormtail}) and Figure~\ref{f:concbnds} consider the
decay with respect to $K$ and the effect of the mean, $\mu_{\alpha_K} = K \mu_{\alpha}$,
becoming more negative as $K$ increases is countered to some extent by the convolution with $\falpha{\alpha}$
broadening the distribution as it evolves with increasing $K$.

This approach also shows that as $K \longrightarrow\infty$ the probability that $x_K$
exceeds any arbitrarily small number goes to zero.  The following lemma states this more
formally.

\begin{lemma}[Lognormal convergence to zero]
\label{l:convtozero}
Assume that $a \sim \rvLN(\mu_a,\sigma^2_a)$, 
$\mu_{\alpha} < 0$, and for simplicity $x_0 = 1$.  For any $\epsilon > 0$,
\[
\lim_{K \longrightarrow \infty} \, \Prob \{ x_K > \epsilon \} \; = 0.
\]
\end{lemma}

\begin{IEEEproof}
We can again write
\begin{IEEEeqnarray*}{rCl}
\Prob \{ x_K > \epsilon \}  &  = &  \Prob \{ \alpha_K \, > \, \ln(\epsilon) \}  \\*[0.5em]
  & = & \left( 1 - \erf{ \frac{\ln(\epsilon) - K\mu_{\alpha}}{\sqrt{2 K \sigma^2_{\alpha}} } } \right).
\end{IEEEeqnarray*}
For all $K > \ln(\epsilon)/\mu_{\alpha}$, the argument of the erf function is positive and 
increasing without bound as a function of $K$.   As
\[
\lim_{x \longrightarrow \infty} \erf{x} = 1,
\]
the result follows. \hfill \IEEEQEDhere
\end{IEEEproof}

It is interesting that this result holds even though $\mean(x_K)$ may be growing to $+\infty$.

\subsection{More general distributions}

The tail probability results above can be generalised to a wider range of distributions and the 
strength of the bound depends upon the assumptions placed on the underlying distribution.  For
a wide range of distributions exponential bounds are still possible, and some examples are provided
below.

A decaying bound is available under the assumptions that $\mu_{\alpha} < 0$ and that the distribution
$\falpha{\alpha}$ has a finite variance, $\sigma_{\alpha}^2$.
These conditions are weaker than those considered for median stability in Theorem~\ref{t:medstab}.
Under the assumption of pairwise independence of the $\alpha_k$ variables, which is satisfied here
by assumption, Cantelli's inequality~\cite{cantelli:1928a} leads to the following bound.

\begin{lemma}
\label{l:cantelliBound}
Assume that $\alpha \sim \falpha{\alpha}$ has a finite mean, $\mu_{\alpha} < 0$, and a finite
variance, $\sigma_{\alpha}^2$.  Then,
\begin{equation}
\Prob\{ x_K \; > \; 1 \} \: \leq \: \frac{1}{1 + K \frac{\mu_{\alpha}^2 \IEEEstrut}
		{\sigma_{\alpha}^2}}. 
\label{e:cantellibound}
\end{equation}
\end{lemma}
 This bound is also illustrated in Figure~\ref{f:concbnds}.  In this general case the distribution
 converges to zero with a $1/K$ rate.  A more general version of Lemma~\ref{l:convtozero} is immediate.

\begin{lemma}
\label{l:generalconvergence}
Assume that $\alpha \sim \falpha{\alpha}$ has a finite mean, $\mu_{\alpha} < 0$, and a finite
variance, $\sigma_{\alpha}^2$.  Then for any $\epsilon > 0$,
\[
\lim_{K \longrightarrow \infty} \, \Prob \{ x_K \, > \, \epsilon \} \; = 0.
\]
\end{lemma}

The Cantelli inequality (Lemma~\ref{l:cantelliBound}) requires the fewest
assumptions on the distribution $\falpha{\alpha}$ and has only a decay rate approximating
$1/K$ for large $K$.  For smaller values of $K$ this bound is actually more accurate than
some of the other bounds.  There exist distributions for which the Cantelli bound is tight and so
in some cases it is not possible to find a better bound.

Tighter bounds are possible if higher moments are known and the next most significant assumption
is that $\alpha \sim \falpha{\alpha}$ comes from a distribution that has
a finite moment generating function within an open interval around zero.   This is equivalent to all
moments of the distribution being bounded.   Distributions not 
satisfying this assumption can be defined as being ``heavy-tailed''.   Note that this assumption is on
the $\alpha$ random variable---the $a \sim \fa{a}$ may be heavy tailed and 
Section~\ref{sb:egheavy} gives a rather extreme example.

The moment generating function is defined as,
\[
\phi_{\alpha}(\lambda) \; := \; \dexpec{ \e{\lambda \alpha} },
\]
and this is assumed to be finite within a region of the origin,
\begin{equation}
\phi_{\alpha}(\alpha) \, < \, \infty, \quad \text{for all }  |\lambda| \leq \beta, \: \beta > 0.
\label{e:mgffinite}
\end{equation}
The moment generating function is used in the calculation of the Chernoff bound
on the tail on the distribution.  In this case,
\begin{multline*}
\Prob \{ \alpha - \mu_{\alpha} \, > \, t \}  \\*[0.5em]
 \leq \; \e{ - \left( \displaystyle \sup_{\lambda \in [0,\beta]}
		\left( \lambda(t - \mu_{\alpha}) - \ln(\phi_{\alpha}(\lambda)) \right)  \right)}.
\end{multline*}
This is not the most general form of the Chernoff bound and other forms give
tighter bounds for low values of $t$.   However the behaviour of the $\fxK{x}$ distribution for large
values of $K$ is the primary concern of this paper and is addressed by the simpler bound given above. 
This can be applied directly to evolution of the $\falpha{\alpha}$ distribution in the following way.

\begin{lemma}
\label{l:chernoff}
Assume that $\alpha \sim \falpha{\alpha}$ has a moment generating function that is
finite over an open interval including zero~(\ref{e:mgffinite}).  Assume also that $\mu_{\alpha} < 0$. Then,
\[
\Prob \{ x_K  \, > \, 1  \}  \: =  \:  \Prob \left\{ \prod_{k=1}^K a_k \, > \, 1 \right\} \: \leq \: \e{ - cK},
\]
where,
\[
c \: = \:  \displaystyle \sup_{\lambda \in [0,\beta]}
		-\lambda \mu_{\alpha} - \ln(\phi_{\alpha}(\lambda)).
\]
\end{lemma}

The proof of Lemma~\ref{l:chernoff} follows immediately from substituting,
\[
\phi_{\alpha_K}(\lambda) \; = \; \phi_{\alpha}(\lambda)^K,
\]
 and $t = 0$ into the Chernoff bound.   So the existence of a finite moment generating function around zero
implies an exponential decay of the distribution of $x_K$ as $K \longrightarrow \infty$.
However, calculating the constant for the exponent requires knowledge of the moment
generating function.

The Chernoff bound (Lemma~\ref{l:chernoff}) is also shown in Figure~\ref{f:concbnds}.
The exponent on this bound is the closest
single exponent bound for the actual tail distribution.  Tighter exponential bounds require sums of exponentials.
This bound can also be tightened by scaling by 0.5.  See the discussion in~\cite{chang:2011a} and references therein for
further details.   However, having $\phi_{\alpha}(\alpha)$ finite in an open interval of the origin uniquely determines
the probability density function and the corresponding cumulative probability density function.  This can then be integrated
numerically to calculate the required probability. 

\subsection{A heavy-tailed example}
\label{sb:egheavy}

\begin{figure*}
\begin{center}
\begin{tabular}{cc}
\includegraphics{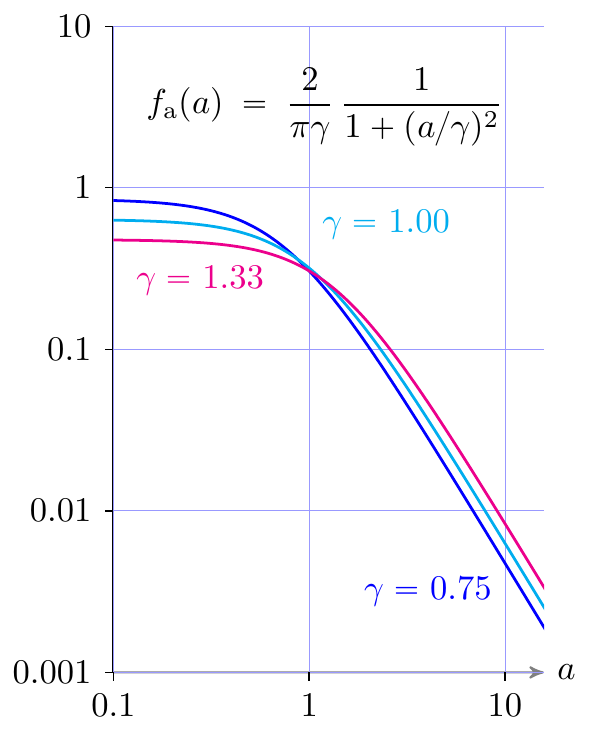} & 
\includegraphics{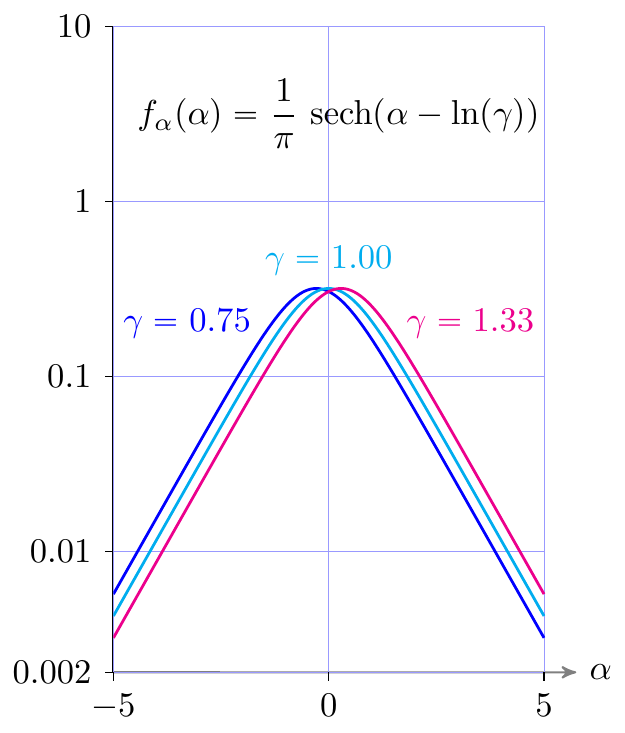}
\end{tabular}
\end{center}
\caption{\label{f:heavydist}  Heavy-tailed example probability distributions. (Left)  $\fa{a}$ distribution on 
a log-log scale.  The linear decay of the tail on the log-log plot shows a power
law characteristic with decay $a^{-2}$.  All moments of the $\fa{a}$ distribution are infinite.
(Right) $\falpha{\alpha}$ distribution on a log-linear scale.  All moments of the $\falpha{\alpha}$ distribution are finite.}
\end{figure*}

It is natural to ask how ``heavy-tailed" the distribution of $\fa{a}$ can be and still lead to median
stability,
\[
\lim_{K \longrightarrow \infty} \median(x_k) \; = \; 0.
\]
To illustrate an extreme case consider the $\fa{a}$ probability distribution to be given by,
\begin{equation}
\label{e:falphadefn}
\fa{a} \: = \: \left\{ \begin{array}{ll}
		\displaystyle \frac{2}{\pi\gamma}\, \frac{1}{1 + (a/\gamma)^2} & \text{for } a\geq 0, \\
		\noalign{\medskip}
		0 & \text{for } a < 0,
		\end{array}
		\right.
\end{equation}
where $\gamma > 0$ is a real-valued parameter.  The left plot in Figure~\ref{f:heavydist} illustrates this distribution
on a log-log scale for three choices of the parameter $\gamma$.    
The distribution is equal to the magnitude of a Cauchy distribution and all of the moments of
this distribution, including the mean, are infinite.   This is also clear from the $a^{-2}$ power law decay in the tail 
shown in Figure~\ref{f:heavydist}(left).    

The calculation of the $\falpha{\alpha}$ distribution is given by~(\ref{e:vdist2alphadist})
and in this case is,
\begin{IEEEeqnarray}{rCl}
\falpha{\alpha} &  =  &  \frac{2}{\pi} \, \frac{ \e{\alpha}/\gamma}{1 + \left(\e{\alpha}/\gamma\right)^2}
     \: = \: \frac{2}{\pi} \, \frac{\e{(\alpha - \ln(\gamma))}} { 1 + \e{2(\alpha - \ln(\gamma))} } \nonumber\\
   \noalign{\smallskip}
    &  = &  \frac{1}{\pi} \, \frac{1} {\cosh(\alpha - \ln(\gamma))} \nonumber \\*[0.5em]
     & = &  \frac{1}{\pi} \, \sech(\alpha - \ln(\gamma)).
\label{e:alphadist}
\end{IEEEeqnarray}

This distribution---without the $\gamma$ parameter---is known in the statistics literature as a hyperbolic secant distribution and has been studied 
for almost 100 years~\cite{fischer:1921a,dodd:1925b,roa:1924a,perks:1932a}.
Most of the main properties of the distribution can be found in~\cite{fischer:2014a}.  The applications of
the hyperbolic secant distribution are not all that common~\cite{ding:2014a,fischer:2014a}. 
There are a range of generalisations to the distribution
with application to specific domains in finance and actuarial statistics; see~\cite{perks:1932a,harkness:1968a}
and~\cite{fischer:2014a}.

Figure~\ref{f:heavydist}(right) shows the probability density function of the $\falpha{\alpha}$ distribution for three choices
of $\gamma$.  The exponential decay of the PDF is clear from the log-linear plot.   All moments of this distribution 
are finite and the moment generating function (for $\gamma = 1$) is,
\begin{equation}
\label{e:heavymgf}
\phi_{\alpha}(\lambda) \: = \: \frac{1}{\cos(\pi \lambda/2)}, \quad |\lambda| < 1.
\end{equation}
The symmetry of~(\ref{e:alphadist}) about $\alpha= \ln(\gamma)$ shows that,
\[
\mu_{\alpha} \: = \: \ln(\gamma).
\]
So by applying Theorem~\ref{t:medstab},
\[
\lim_{K\longrightarrow\infty} \median(x_K) \, = \, 0 \quad \iff \quad \gamma < 1.
\]
Note however that for all $\gamma > 0$, 
\[
\dexpec{x_K } \; = \; \infty
\]
for all $K$.  This is an extreme example of an unstable mean.

\begin{figure}
\centering
\includegraphics[width=\columnwidth]{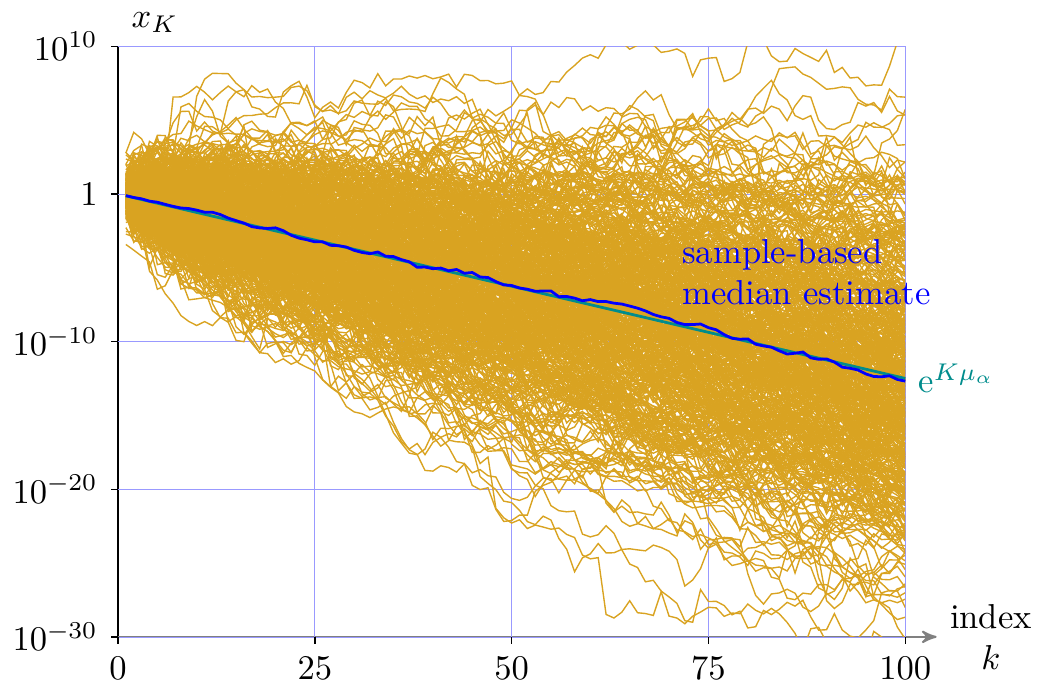}
\caption{\label{f:logsechsims}500 sample trajectory simulations for $\gamma=0.75$.  The median of 
the $\zeta_K$ distribution is compared with a sample estimate of the median.}
\end{figure}

The symmetry of the $\falpha{\alpha}$ distribution implies that the median and mean are equal
and the median of the state $x_K$ can therefore be given analytically,
\begin{IEEEeqnarray*}{rCl}
\median(x_K) &  = &  \e{\median(\zeta_K)} \\
  		&  = &  \e{\mean(\zeta_K)} \: = \: \e{K\ln(\gamma)} \: = \: \gamma^K.
\end{IEEEeqnarray*}
For illustration, and comparison with Figure~\ref{f:lognormcase3}, 500 random trajectories of $\zeta_K = \ln(x_K)$,
for $K = 1,\dots,100$ are shown in Figure~\ref{f:logsechsims}.   The predicted evolution of the median of 
$\zeta_K$ is compared to a sample-based estimate and found to be accurate.  As a result of the very heavy nature of
the $\fa{a}$ distribution, the range of the $\zeta_K$ trajectories in Figure~\ref{f:logsechsims} is much greater than in the
normal/lognormal case shown in Figure~\ref{f:lognormcase3}.

\begin{figure}
\includegraphics[width=\columnwidth]{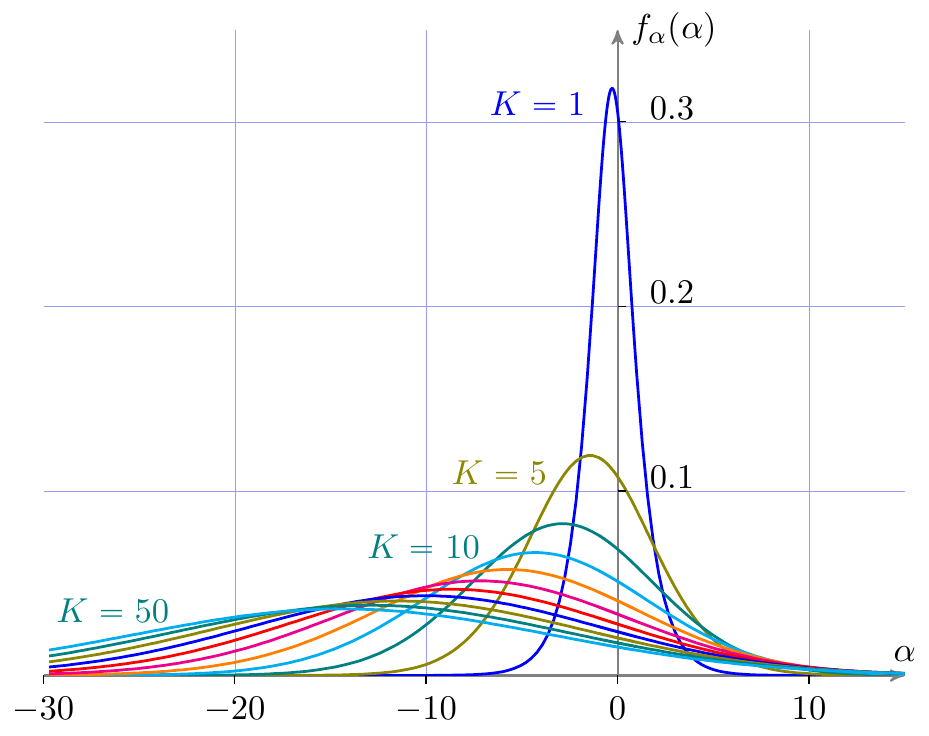}
\caption{\label{f:zetaKdist} Evolution of the probability distribution function of $\zeta_K$ for
$K = 1,5,\dots,50$.  The initial ($K=1$) distribution is that shown in Figure~\ref{f:heavydist}(right) for 
$\gamma = 0.75$.}
\end{figure}

The behaviour of the distribution of $\zeta_K$, as $K\longrightarrow\infty$ is given
by the distribution of the $K$-fold sum of random variables, $\alpha_k$, each drawn
from the $\falpha{\alpha}$ distribution.
So the distribution of $\alpha_K$ is the $K$-fold convolution of $\falpha{\alpha}$ and this
has been numerically calculated (using the Chebfun \Matlab\ Toolbox~\cite{chebfun:2014})
in Figure~\ref{f:zetaKdist}.  The characteristics of a sum of hyperbolic secant random variables
were first studied in~\cite{baten:1934a}.

\begin{figure}
\centering
\includegraphics[width=\columnwidth]{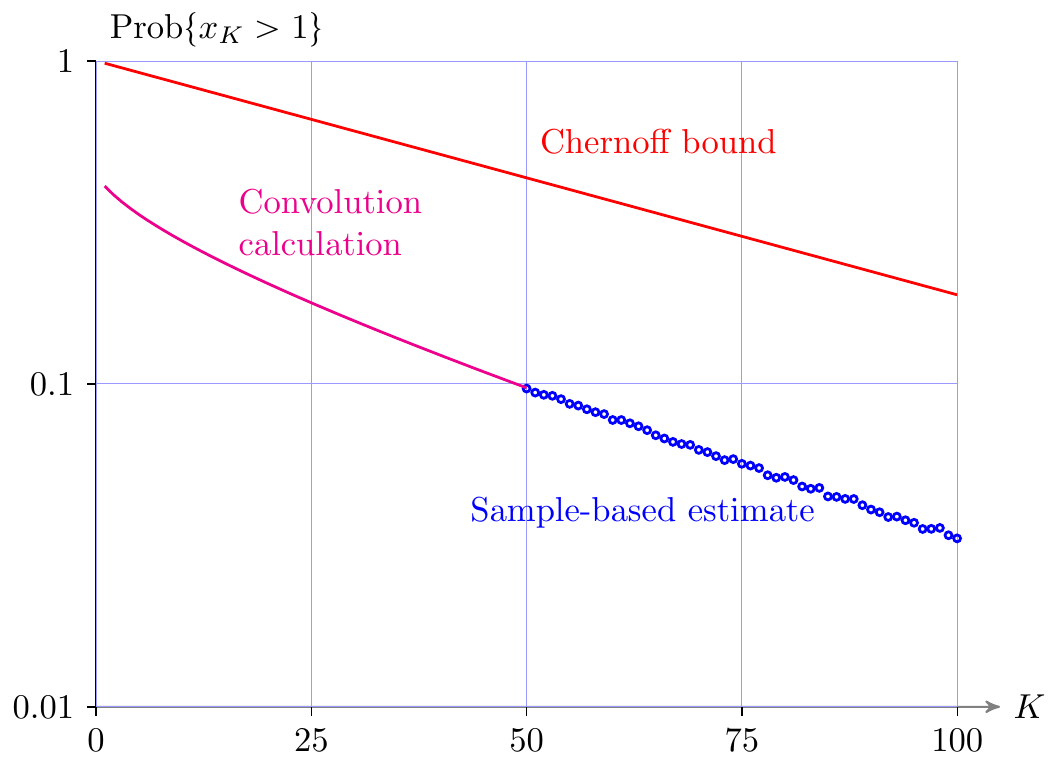}
\caption{\label{f:heavyaK}
Comparison between the Chernoff bound (red) given in~(\ref{e:heavycbnd}),
(\ref{e:heavycfactor}) and (\ref{e:lopt}) 
and $\Prob\{x_K > 1\}$ as a function of $K$ for $\gamma = 0.75$.  For $K$ up to 50 the numerical
evaluation (via Chebfun~\protect\cite{chebfun:2014})
of tail of the $K$-fold convolution of $\falpha{\alpha}$
gives the actual probability and is shown (magenta).  For $K > 50$ (where numerical limitations prevent the numerical
probability calculation) the probability is estimated from 100,000 sample trajectories (blue circles).}
\end{figure}

As the moment generating function for $\falpha{\alpha}$ in~(\ref{e:heavymgf}) is finite in a range around zero
the probability that  $x_K > 1$ decays to zero exponentially as $K\longrightarrow\infty$.  The bound can be 
calculated from the moment generating function in~(\ref{e:heavymgf}) and Lemma~\ref{l:chernoff}.
\begin{equation}
\Prob \left\{ \prod_{k=1}^K x_k \, > \, 1 \right\}  \; \leq \; \e{-cK},
\label{e:heavycbnd}
\end{equation}
where,
\begin{equation}
c \; = \; -\lambda^* \ln(\gamma) + \ln\left( \cos (\pi \lambda^*/2 ) \right),
\label{e:heavycfactor}
\end{equation}
and 
\begin{equation}
\lambda^* \; = \; \frac{2}{\pi} \arctan \left( \frac{-2 \ln(\gamma)}{\pi} \right).
\label{e:lopt}
\end{equation}
Figure~\ref{f:heavyaK} shows this bound.   The actual probably can be calculated numerically
from the distributions in Figure~\ref{f:zetaKdist}, and estimated from samples in the simulation in Figure~\ref{f:logsechsims}.
Both of these comparisons are made and indicate that the exponent in the Chernoff bound is tight but the bound itself could
be divided by a factor of at least two.  

This is an extreme example and it is interesting to put it into the context of a simple investment finance problem.  
Consider the accumulated return on an investment with an independent, identically distributed, random rate of
return at every time step.  In this context $x_0$ is the initial
investment, $\fa{a}$ is the probability distribution of the rate of return at each time step, and $x_K$ is the investment value after
$K$ time steps.   Then this example is a case where the expected rate of return is infinite for each of the 
time steps, and yet the probability of making a profit after $K$ time steps decays exponentially to zero as $K$ increases.

\section{Discussion}

Our goal in this paper has been to precisely specify and illustrate the conditions 
for the stability of the median, mean and variance in discrete-time stochastic
feedback settings.   The discrete-time setting enables a far wider range of 
distributions to be considered than is possible in the continuous-time case, and
is at the same time relevant to a wide range of problems.   The focus on the scalar
variable case is of course much more restrictive and has allowed precise statements
to be made about the probability of the distributions of solutions to the difference
equations.   This is particularly true for the median statistics.

The differences between the stability conditions arise because of the
heavy-tailed nature of the resulting distributions.  This allows the phenomenon
of the mean growing exponentially while the distribution converges exponentially
to zero to arise.   Note that the stochastic component of the system need not 
be heavy-tailed for this to be observed;  it suffices that the effect of the stochastic 
component is integrated via a feedback interconnection with a dynamical system. 

The variance stability condition given here is a simple case of the more widely
known mean-square stability criterion dating from the 1970s~\cite{willems:1973a}.   
This condition has the advantage that it is also exact for the multi-variable case.
However, it is acknowledged that mean-square stability is a strong form of
stability~\cite[p.~136]{boyd:1994a}.  This results in this paper emphasise this point,
particularly in comparison to median stability.

The conditions for median, mean, and variance stability are different and it is a
natural question to consider which is more appropriate for use in any particular 
problem.    Very different answers can arise from the exact statement of the problem
and can easily lead to interpretations which are---at least from a cursory point of 
view---in contradiction.   For example, in an investment problem there are a relatively
wide range of circumstances in which a return-on-investment will have an expected
value greater than one, and consequently the expected profit grows with time, and yet
in which the probability of making any profit at all decays to zero.   The equivalent
conditions in a population 
dynamics context would indicate that a survival rate may be greater than one and yet
the probability of extinction is also going to one.

These apparent paradoxes illustrate that in stochastic feedback situations seemingly similar
questions may have widely divergent answers and it is important to pose the correct
measure of stability in the problem formulation and the analysis.   The increasing use of
interconnected feedback networks, and particularly those where online data-based updating
leads to stochasticity in the feedback components, requires that we make a careful
choice of analysis criteria and design methods.

\section{Acknowledgements}    

The authors would like to thank Tryphon Georgiou, Sean Meyn, Vikram Krishnamurthy,
and Subhrakanti Dey for useful discussions on this topic.

\bibliographystyle{IEEEtran}

\end{document}